\documentclass[12pt,twocolumn,numberedappendix]{openjournal}

\usepackage{xcolor}
\usepackage{textgreek}
\usepackage[utf8]{inputenc}
\usepackage[english]{babel}
\usepackage{hyperref}
\hypersetup{
    colorlinks=true,
    linkcolor=blue,
    filecolor=blue,      
    urlcolor=red,
    citecolor=blue,
}
\usepackage{color,colortbl}
\usepackage{tensind}
\tensordelimiter{?}
\DeclareGraphicsExtensions{.bmp,.png,.jpg,.pdf}
\usepackage{verbatim}
\usepackage[normalem]{ulem}
\usepackage{orcidlink}
\usepackage{soul}

\graphicspath{ {./figs/} }

\usepackage{newtxtext,newtxmath}
\usepackage[T1]{fontenc}
\DeclareRobustCommand{\VAN}[3]{#2}
\let\VANthebibliography\thebibliography
\def\thebibliography{\DeclareRobustCommand{\VAN}[3]{##3}\VANthebibliography}
\usepackage{graphicx}	
\usepackage{amsmath}	

\begin{document}

\title[Cosmic Dragons]{Cosmic Dragons: A Two-Component Mixture Model of COSMOS Galaxies}

\author{\vspace{-1.5cm}William K. Black\orcidlink{0000-0003-4811-7913}}
\email{wkblack@umich.edu}

\author{\hspace{.2cm}August E. Evrard\orcidlink{0000-0002-4876-956X}}
\affiliation{Department of Physics and Leinweber Center for Theoretical Physics, University of Michigan}

\begin{abstract}
  Using the photometric population prediction method {\bf Red Dragon}, we characterize the Red Sequence (RS) and Blue Cloud (BC) of DES galaxies in the COSMOS field. Red Dragon (RD) uses a redshift-evolving, error-corrected Gaussian mixture model to detail the distribution of photometric colors, smoothly parameterizing the two populations with relative weights, mean colors, intrinsic scatters, and inter-color correlations. 
  This resulting fit of RS and BC yields RS membership probabilities $P_{\rm RS}$ for each galaxy. Even when training on only DES main bands $griz$, RD selects the quiescent population (defined as galaxies with $\lg {\rm sSFR \cdot yr} < -11$) with $\gtrsim 90\%$ balanced accuracy out to $z=2$; augmenting with extended photometry from VIRCAM improves this accuracy to $\sim 95\%$ out to $z=3$. 
  We measure redshift evolution of sSFR and galactic age in several stellar mass bins, finding that the BC is consistently more star-forming (by $\gtrsim 1~{\rm dex}$) and typically younger (by $\gtrsim 1~{\rm Gyr}$) than the RS (up to $z \sim 1.4$). This characterization of both RS and BC as functions of redshift and stellar mass improves our understanding of both populations and opens the door to more precise galaxy population characterization in future deep optical and IR systems. 
\end{abstract}


\maketitle


\section{Introduction}
The galaxy population consists of two main flavors: 
  a ``blue cloud'' (BC) of actively star-forming galaxies (also called the galactic main sequence) 
  and 
  a ``red sequence'' (RS) of passively evolving (``red and dead'') galaxies which have largely ceased star formation \citep{Strateva+01, Bell+04}. 
  Between these two populations lies the so-called ``green valley'' (GV). 
The BC typically is made of young, spiral galaxies commonly found in low-density environments. 
In contrast, RS galaxies are typically massive, old, bright ellipticals, and are usually found in high-density environments---especially the interiors of galaxy groups and clusters. 
Only about 6\% of galaxies are either red spirals or blue ellipticals \citep{Schawinski+14}; the lion's share of galaxies fall into this bimodal distribution. 

If a galaxy has no cold gas which can condense into new stars, star formation ceases, and the galaxy is considered \emph{quenched}. Various quenching mechanisms dominate at different galactic mass ranges and at different redshifts \citep[see e.g. Figure~15 of][]{Peng+10}, either heating up or removing the galactic gas. 
Though the exact physics behind the quenching of satellite galaxies is unknown, several processes play crucial roles \citep{Somerville+15, Alberts+22}. 
Galaxies with decimal log stellar mass $\log_{10} M/M_{\odot} \gtrsim 10.5$ have supermassive black holes which blow gas out of the galaxy, preventing formation of new stars. 
Especially at early times, merging events play crucial roles in galaxy quenching---disrupting, heating, and ejecting gas from galaxies \citep{LaceyCole1993mergerRates, DonahueVoit2022MassiveGalReview}. 

In addition to initiating merging, gravity attracts galaxies towards denser regions, which quench galaxies. 
As a galaxy barrels through the hot gas of a cluster or group, the resulting ram pressure heats and strips away gas. Close interactions between galaxies (or groups of galaxies) cause tidal stripping, pulling off outer layers of cold gas from the galaxy. 
These stripping processes make galaxy clusters hotbeds for the creation of quiescent galaxies (galaxies which have largely ceased star formation). As clusters are the endpoint of large-scale gravitational collapse, they also serve as the terminal location of such galaxies. 
Furthermore, in a Gaussian random field, galaxy clusters host some of the earliest forming galaxies \citep{Springel+05}, implying that the central galaxies of clusters are likely to host massive and old stellar populations \citep{Collins2009earlyMassiveGals}.
Therefore, as compared to other locations of the cosmic web, galaxy clusters tend to have higher red fractions \citep{DeLucia+04, Tanaka+05, Hansen+09, Pandey+20}. 
Red fraction helps determine whether quenching occurred due to mass or environment \citep{Tanaka+05, Baldry+06, Peng+10}.

Though no absolute boundary exists between RS and BC, various means exist to classify galaxies in a strongly bimodal fashion. 
For example, the S\'ersic index, H$\alpha$ width, and features of the circular velocity can all help distinguish the two populations \citep{Krywult+17, Kalinova+22}. 
The most striking difference in the spectra of RS and BC galaxies in the optical regime 
is ``the $4000~\Angstrom$ break'' \citep{Balogh+99, Hathi+08, Kriek+11, Kim+18}, a feature primarily driven by differences in a galaxy's current star formation rate. 
Dust, metallicity, age, and other factors imprint additional features that can help distinguish RS \& BC components 
\citep{Fabbianio1989, Worthey_1994, Blanton_Moustakas_2009, Symeonidis2022QSOSFR}. Galaxy population models can therefore make good use of spectral features other than the 4000~$\Angstrom$ break alone to deduce component membership.

Viewing the distribution of photometric color about the 4000~$\Angstrom$ break (as well as in other wavelength regions), bright RS galaxies tend to appear redder and tighter-clustered than the relatively blue and loosely-clustered BC galaxies (hence their names of ``red sequence'' and ``blue cloud'' respectively). This observed bimodality arises for two primary reasons: 
  First, specific star formation rates (sSFR; star formation rate per galactic stellar mass) follow a skew-lognormal distribution, peaking at high star formation ($\lg {\rm sSFR \cdot yr} \sim -10$ near $z=0$) for the galactic main sequence (BC galaxies) with a long tail towards lower star formation for the RS \citep{Wetzel+12, Eales+18, Leja+22}. 
  Second, at low sSFR values ($\lg {\rm sSFR \cdot yr} \lesssim -11.3$), the scatter in color at fixed sSFR decreases significantly, such that low-sSFR galaxies all tend towards the same location in color space \citep{Eales+17}. 
These two effects then conspire to result in a dual Gaussian distribution of galaxies in photometric color space \citep{Strateva+01, Bell+04, Baldry+04, Williams+09, Hao+09, Krywult+17}. 
We therefore can use Gaussian mixtures to characterize these populations in multi-color space.

Gaussian mixture models (GMMs) can classify photometric colors into groups, where each component $\alpha$ has weight $w_\alpha$, mean color vector $\vec \mu_\alpha$, and color covariance matrix $\Sigma_\alpha$ (including color scatters $\sigma$ and correlations between colors $\rho$ for multi-color models). 
GMMs effectively differentiate between galaxy populations in color space, offering a data-driven classification method that avoids the need to select and justify specific cutoff criteria \citep{Baldry+04, Krywult+17, Ardila+18, Siudek+18, Gould+23, Dogruel+23}. 
For a two-component model of the RS and BC, this then entails the total parameter set $\{ f_{\rm RS}, \vec\mu_{\rm RS}, \vec\mu_{\rm BC}, \Sigma_{\rm RS}, \Sigma_{\rm BC} \}$, where red fraction $f_{\rm RS} \equiv w_{\rm RS} / (w_{\rm RS} + w_{\rm BC})$. 
Each of these parameters evolve over redshift and stellar mass for both RS and BC populations \citep{Baldry+04, Balogh+04, Baldry+06, Ruhland+09}. 
As mentioned previously, red fraction additionally depends on local overdensity, but dependence of the other parameters on local density have yet to be significantly measured separately from their dependence on stellar mass (high-mass galaxies are born and bred in denser environments, so the two factors correlate).

In this paper, we characterize GMM fit parameters for galaxies in the COSMOS field using the Red Dragon algorithm. Parameterization of the RS and BC then yields probabilistic component classification for an individual galaxy. 
  In \S\ref{sec:RD_intro}, we describe the algorithm and discuss expected dependence of variables with stellar mass and redshift. 
  In \S\ref{sec:data}, we detail the galactic data used in this analysis, including its mass completeness, redshift limits, and color selection. 
  In \S\ref{sec:RDii_Results}, we examine stellar mass and redshift dependence of the two color populations (\S\ref{sec:fit}; additional details in Appendix~\ref{apx:other_colors}).  Using published ages and star formation rates for COSMOS galaxies, we explore statistics of these measures for RS \& BC galaxy sub-populations  (\S\ref{sec:P_red}). 
  In \S\ref{sec:RDii_Discussion}, we discuss quiescent population selection accuracy, color scatter as a function of rest frame wavelength, and the optimal component count with which to characterize the photometric galaxy populations. 
  We summarize our core findings in \S\ref{sec:RDii_Conclusion}.

\section{Red Dragon Algorithm} \label{sec:RD_intro}

\defcitealias{Black+22}{B22}

To parameterize the RS and BC, we use the galaxy population modeling tool {\bf Red Dragon} \citep[RD,][]{Black+22}.
RD uses redshift-evolving, error-corrected Gaussian mixtures in photometric multi-color space to characterize the galaxy population as a superposition of RS and BC components (as well as optionally additional components). 
In this section, we briefly review the algorithm and discuss previous fit parameterizations from observed and synthetic data. 
The resulting parameterization of RS and BC not only gives phenomenological descriptions of the evolution of each population but also allows for probabilistic classification for individual galaxies.

\vspace{.5cm}

\subsection{Algorithm overview} \label{sec:RD_algorithm}

Red Dragon uses an evolving, error-corrected Gaussian Mixture Model (GMM) in the multi-dimensional space of photometric colors to characterize galaxy populations. 
For a given component (e.g. RS or BC) $\alpha$, the likelihood of the parameterization $\theta_\alpha$ for a galaxy characterized by $x_j$ (defined below) is 
\begin{equation}
\begin{aligned} \label{eqn:L_ij}
  \mathcal{L}_\alpha ( \theta_\alpha \big| x_j) = \, 
  & \frac{w_\alpha}{\sqrt{(2\pi)^{D}} \left| \Sigma_\alpha + \Delta_j \right|} \\
    & \times \exp 
      \left[ 
        -\frac12 
        (\vec c_j - \vec \mu_\alpha)^\mathrm{T} 
        (\Sigma_\alpha + \Delta_j)^{-1}
        (\vec c_j - \vec\mu_\alpha) 
      \right] ,
\end{aligned}
\end{equation}
where the parameter set $\theta_\alpha$ is composed of components
\begin{itemize}
  \item weight $w_\alpha$ (sum of all weights constrained to unity), 
  \item mean color vector $\vec \mu_\alpha$ (length $D$), and
  \item color covariance matrix $\Sigma_\alpha$ (size $D \times D$); 
\end{itemize}
and each galaxy $x_j$ has inputs of its
\begin{itemize}
  \item measured photometric colors $\vec c_j$ and
  \item noise covariance matrix $\Delta_j$ (encoding uncertainties). 
\end{itemize}
The algorithm \texttt{pyGMMis} \citep{Melchior_Goulding_2018} performs this parameter estimation using ``extreme deconvolution'' \citep{Bovy+11} to estimate the error-free (intrinsic) distribution of colors. The method provides both intrinsic scatter as well as intrinsic correlations, accounting for the stretching induced by uncertainties on the observed (error-included) magnitudes.

The parameterizations $\{\theta_\alpha\}$ for each of the $K$ components can then be optimized by finding the maximum likelihood parameterization for all $N_{\rm gal}$ galaxies with data $\{ x_j \}$:
\begin{equation}
  \mathcal{L}(\{\theta_\alpha\} \big| \{x_j\}) = \prod_{j=1}^{N_{\rm gal}} \sum_{\alpha=1}^K \mathcal{L}_\alpha ( \theta_\alpha \big| x_j).
\end{equation} 
With bootstrap resampling of the input galaxy population $\{ x_j \}$, one can then obtain uncertainties on all fit parameters.

The Red Dragon algorithm measures these parameterizations across redshift (given a sufficiently thin redshift bin Red Dragon can run across stellar mass or some other variable). \href{https://github.com/afarahi/kllr/tree/master/kllr} {KLLR} \citep{Farahi+18,Farahi+22} then smoothly interpolates these parameterizations with a Gaussian kernel width of $\sigma_z = .05$, yielding a continuous characterization of RS, BC, and optionally additional components. This yields a phenomenological description of the color-space evolution of each population.

With fit characterization in hand, one can then predict for a galaxy $x_j$ at redshift $z_j$ to which component it more likely belongs. The probability of it belonging to component $\alpha$ is
\begin{equation} \label{eqn:P_alpha}
  P_\alpha(x_j) = \frac{ \mathcal{L}_\alpha (\theta_\alpha \big| x_j) } { \sum_\beta \mathcal{L}_\beta ( \theta_\beta \big| x_j)}. 
\end{equation}
For the standard two-component RS \& BC model of galaxy populations, this then yields $P_{\rm RS} = \mathcal{L}_{\rm RS} / (\mathcal{L}_{\rm RS} + \mathcal{L}_{\rm BC})$ as the probability of the galaxy belongs to the RS. Red Dragon thus calculates probabilistic membership classification for individual galaxies. 

Code for Red Dragon is freely available \href{https://bitbucket.org/wkblack/red-dragon-gamma/}{on BitBucket}.\footnote{\href{https://bitbucket.org/wkblack/red-dragon-gamma/}{bitbucket.org/wkblack/red-dragon-gamma}}

\subsection{Expected parameter dependence}
\label{sec:previous_GMM}

Gaussian parameterization of the RS and BC (red fraction, mean color, and color scatter) depends on redshift, galactic stellar mass, and local overdensity \citep{Balogh+04}. 
\citet{Baldry+04} quantified for a rest-frame sample of low-$z$ galaxies the evolution of red fraction, mean $u-r$ color (a proxy of star formation rate), and color scatter as a function of $M_r$ magnitude (a proxy of galactic stellar mass). 
They fit color and scatter as linear functions with a hyperbolic tangent transitioning the fit from one vertical intercept to another, resulting in a slanted sigmoid function. 
Colors redden monotonically with increased log stellar mass. 
In particular, after an inflection point at $\lg M_\star / M_{\odot} \sim 10.35$, colors jump from a bluer trend to a redder. 
Thus linear extrapolation from high-masses ($\lg M_\star / M_{\odot} \gtrsim 11$) down will predict significantly redder galaxies at lower masses than observed. 
In contrast to mean colors, while color scatters generally increase with stellar mass, they only do so monotonically so for the RS. At $\lg M_\star / M_\odot \sim 10.0$, BC scatter increases momentarily, then continues its decreasing trend with increasing stellar mass, resulting in minimal net trend. 
While these measurements only include a single rest-frame color, more are available using DES $griz$ photometry.

Using data from DES Y3, {\sc redMaPPer} \citep[RM;][]{Rykoff+14} characterizes bright\footnote{
  $L > 0.2 \, L_{*,z}(z)$, 
  where $L_{*,z}(z)$ is the characteristic $z$-band luminosity for a galaxy at a given redshift $z$
} members of the RS in a multi-color $\times$ magnitude space, evolving a Gaussian parameterization of the RS across redshift. 
RM allows mean RS colors to drift linearly with magnitude (such that brighter galaxies were redder), allowing that slope to evolve with redshift. 
Such linear modeling matches the high-mass characterizations of \citet{Baldry+04}, as these galaxies were generally heavier than the sigmoid transition mass mentioned above. 
%
While RM accounts for uncertainties in estimating intrinsic scatter, it does not account for uncertainties in measurements of correlations; in regions where color uncertainties exceeded intrinsic scatter (which was nearly always, since it trained on a photometric dataset), inter-color correlations more so reflect correlations between {\it uncertainties} rather than reflecting {\it intrinsic} (error-free) correlation between RS colors, as RD measures. We therefore do not compare their correlations to our own. 
Furthermore, as RM only considers the RS, it lacks red fraction and BC outputs. 
For these reasons, in future comparisons to this work, we will only show the RS mean color and scatter as estimated by this model.


\subsection{Fit Results from B22}
\label{sec:RD_results}

Previous Red Dragon fitting of RS and BC done by \citet[][hereafter \citetalias{Black+22}]{Black+22} used bright\footnote{
  $L > 0.2 \, L_{*,i}(z)$, where $L_{*,i}(z)$ is the $i$-band characteristic magnitude at a given redshift $z$
} galaxies from the Sloan Digital Sky Survey \citep[SDSS;][]{Szalay+02} and a DES-like synthetic galaxy catalog \citep[Buzzard;][]{DeRose+19,DeRose+21,Wechsler+22}. 
The $ugriz$ SDSS sample focused on low redshifts $z = .1 \pm .005$ and intermediate redshifts $z \in (.3, .5)$; in contrast, the $griz$ Buzzard sample covered a wider redshift range $z \in [.05, .84]$. Several key results follow.

In both SDSS samples, RD selected the quiescent population, defined as $\lg ({\rm sSFR \cdot yr}) < -11 + z$, with $\gtrsim 92\%$ accuracy.\footnote{
  In the low-$z$ sample, increasing the threshold to $\lg {\rm sSFR \cdot yr} < -10.7$ decreased selection accuracy to $\sim 89\%$ while decreasing the threshold to $< -11.3$ increased selection accuracy to $\sim 94\%$. 
}
A three-component fit was found to best explain the data; the third component (beyond RS and BC) had lower weight ($w<10\%$), higher scatter (roughly twice that of the BC), and had low (small positive or consistent with null) inter-color correlations (often the lowest). This suggests the third component captured `noise', i.e. galaxies that didn't fit well in either component. 
In the mid-$z$ SDSS sample, we investigated RD selection accuracy across a redshift transition of the $4000~\Angstrom$ break; we found either similar or superior selection across the entire redshift span compared to single-color selection, with $>6 \sigma$ superiority at the point of transition.

In Buzzard, the red fraction $f_{\rm RS}$ decreased near linearly with redshift. Mean colors agreed well with RM fits (E. Rykoff 2022, private communication), though at $z \gtrsim .7$ RM measured somewhat bluer colors than RD. 
Measurements of scatter roughly agreed, with RM measuring scatters $\sim 50\%$ smaller than RD. 

RD made the first measurement of intrinsic inter-color correlations: correlations between pairs of photometric colors within a population. 
In both SDSS as well as Buzzard, Red Dragon widely measured inter-color correlations of $\rho_{\rm RS} < \rho_{\rm BC} < 95\%$. 
In Buzzard, correlations roughly follow $\rho_{\rm BC} \sim .95 - .2 z$ and $\rho_{\rm RS} \sim .8 - .5 z$, though near $4000~\Angstrom$ break redshift transitions some correlations dipped towards zero. 
These results show considerably lower RS inter-color correlations than expected.

\section{Data} \label{sec:data}
We use galaxies from one of the DES deep fields known as the COSMOS patch. Utilizing photometry, this sample estimates redshifts, stellar masses, star formation rates, and ages for galaxies. In this section, we detail the COSMOS dataset (\S\ref{sec:COSMOS2015}) and the particular selection of galaxies we utilize in this paper. 
We ensure mass completeness for each of the three main samples (\S\ref{sec:cuts/mass}) then focus on redshift bins which contain a sufficient quantity of quiescent galaxies (\S\ref{sec:cuts/z}). We also remove a small number of extreme outliers in color space (\S\ref{sec:cuts/col}). 
Red Dragon then fits this resulting selection of photometric colors with a dual Gaussian mixture as a function of redshift for each bin of our three mass decades.


\subsection{The COSMOS2015 Catalog} \label{sec:COSMOS2015}

We employ data from the \href{https://cosmos.astro.caltech.edu/page/public}{COSMOS} (Cosmic Evolution Survey) field \citep{Yoshiaki+05,Scoville+07}, chosen for its high Galactic latitude, minimal foreground contamination, 
and optimal observability. We focus on a portion of the field centered at J2000 $({\rm RA}, {\rm Dec.}) = (150.1166, 2.2058)$, covering an area of 1.24 sq. deg. 

The COSMOS field has received spectrum-wide exposure, with observations ranging from X-ray to radio wavelengths \citep{Schinnerer+04,Hasinger+07}. We focus on the COSMOS2015 catalog: observations from the COSMOS DES (Dark Energy Survey) Deep Field \citep{Laigle+16,Laigle+18}. 
Data come from the selection of \citet{Hartley+22}, combining data from both the Dark Energy Camera \citep[DECam;][]{Flaugher+15}, which supplies bands $ugriz$, as well as from the VISTA InfraRed CAMera \citep[VIRCAM;][]{Dalton+06,Emerson+06}, which supplies bands $JHK_{\rm s}$ \citep[as part of the \href{https://ultravista.org/}{UltraVISTA} survey; see][]{McCracken+12,Caputi+15}.

\begin{figure}\centering
  \includegraphics [width=\linewidth] {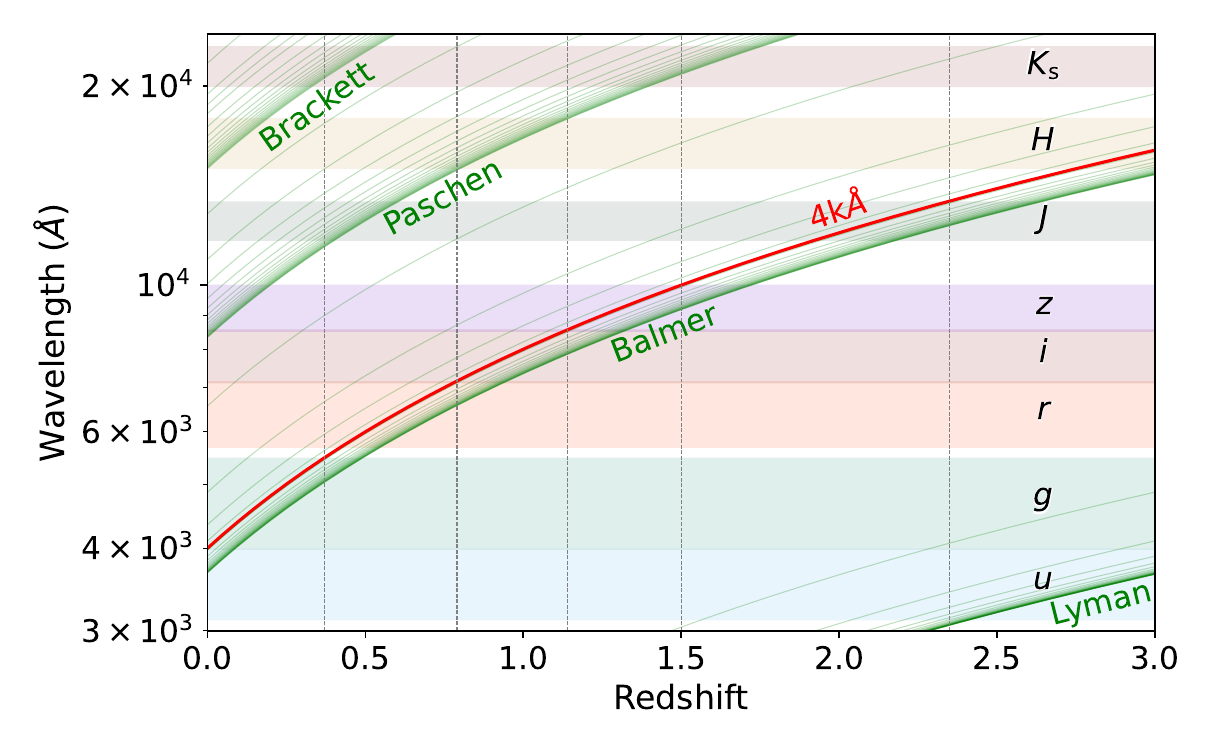}
  \vspace{-16 truept} 
  \caption[Redshift drift of galactic spectral features]{
    Redshift drift of several galactic spectral features as observed by photometric bands from DECam and VIRCAM. 
    Hydrogen spectral lines are shown in green. 
    In red is the $4000~\Angstrom$ break, the strongest distinguishing feature between RS and BC spectra; vertical grey lines indicate redshifts at which the break exits each band. 
  }
  \label{fig:z_transitions}
\end{figure}

Figure~\ref{fig:z_transitions} displays these bands, showing both their observing wavelengths (at $z=0$) as well what galactic spectral features they measure at higher redshifts. DECam $ugriz$ bands span $\lambda \sim~3100$ to $10,000~\Angstrom$ while VIRCAM $JHK_{\rm s}$ bands span $\lambda \sim~11,700$ to $23,000~\Angstrom$. Central wavelengths and widths for each band are given in Table~\ref{tab:break_locations}, along with the redshifts at which the $4000~\Angstrom$ break enters and exits each band (shown as vertical grey lines above).

We employ physical quantities derived from ultraviolet to mid-infrared photometry are from the COSMOS\-2015 catalog \citep{Laigle+16}. 
{\sc LePhare} \citep[PHotometric Analysis for Redshift Estimations;][]{Arnouts+02,Ilbert+06} uses a suite of \citet{BC03} spectrum templates to model expected redshifts (median redshift uncertainty $\sigma_z/(1+z) = .036$ in our mass ranges). 
This model used exponentially declining star formation histories (SFHs) as well as delayed SFHs ($\tau^{-2} \, t \, {\rm e}^{-t/\tau}$) at ages $\tau = 1$ and $\tau = 3~{\rm Gyr}$ \citep{Ilbert+15}. 
The SED fitting provides estimates of stellar mass (median uncertainties $<.1~{\rm dex}$ in each mass range), specific star formation rates (median uncertainties $\lesssim .25~{\rm dex}$ in each mass range), and light-weighted galactic ages (no uncertainties provided\footnote{
  Estimated ages take on 41 distinct values in our data, spanning from 50~Myr to 12~Gyr, indicating distinct BC03 template models. Spacing between age values ranges from 20~Myr to 1~Gyr. 
}). 

\subsection{Stellar mass completeness}
\label{sec:cuts/mass}

\citet[][Table~2]{Ilbert+13}, gives the minimum masses at which the RS \& BC are complete for several redshift bins in the COSMOS field. As the BC minimum masses lie below those of the RS, we use the former. The minimum mass at a given redshift is well fit by 
\begin{equation} \label{eqn:a_min}
  \lg M_{\star, {\rm min}}(z) / M_\odot = 11.4 - \frac{5.2}{1+z}. 
\end{equation}
This equation determines the maximum redshift to which we can extend while remaining complete down to a certain mass: 
lowering the mass completeness requirement limits a sample lower redshifts.

\begin{figure}\centering
  \includegraphics [width=\linewidth] {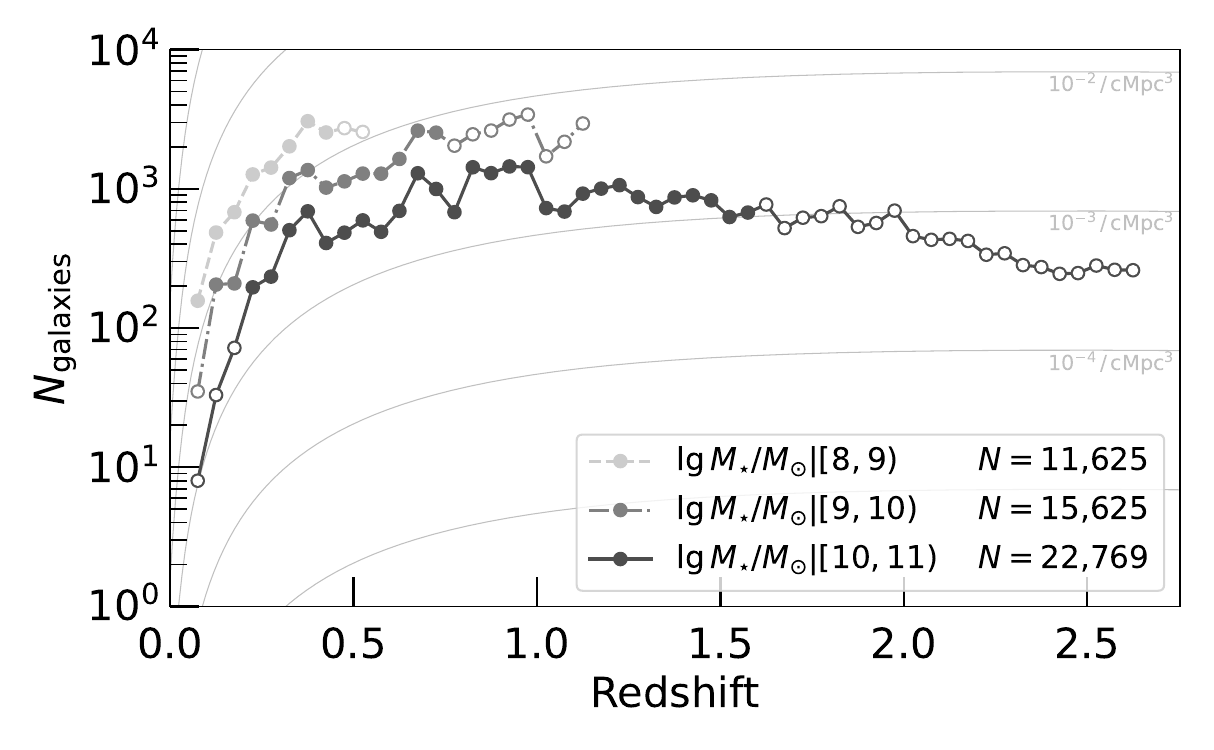}
  \vspace{-16 truept} 
  \caption[Galaxy count per redshift and mass bin]{
    For each decadal stellar mass bin used in this analysis, we show the count of galaxies in each $\Delta z = 0.05$ redshift bin (width used for RD fitting). 
    The legend includes both log stellar mass ranges as well as galaxy count $N$ of each sub-sample used for core results.  
    Open circles mark bins with insufficient quiescent galaxies (hidden in the main results that follow). 
    Grey lines mark constant comoving number densities per redshift bin. 
  }
  \label{fig:Ngal_z}
\end{figure}

Low-mass galaxies are more abundant than high-mass galaxies, yet a lower mass threshold decreases the redshift range; these two factors compete to yield a peak number of galaxies when using a mass completeness limit of $\lg M_{\star, {\rm complete}}/M_\odot \sim 9.14$ (extending out to $z = 0.56$, with a total galaxy count of $N \sim 54,000$). 
Focusing about this value, for our analysis we use three stellar mass samples, each a decade wide, spanning the range $\log_{10} M_\star / M_\odot \in [8, 11)$. Their minimum masses of $\log_{10} M_\star / M_\odot = \{ 8, \, 9, 10 \}$ require maximum redshifts of $z_{\max} = \{.55, \, 1.15, \, 2.65\}$ respectively to ensure mass completeness. 

Figure~\ref{fig:Ngal_z} shows the resulting number counts as a function of redshift for each of these three mass decades. Counts per redshift shell rise as more volume is enclosed (until $z \sim .925$), then gradually fall with redshift as the same $\Delta z$ encloses progressively less volume. As expected from the luminosity function for mass-complete samples, we see significantly more low-mass galaxies than high-mass galaxies at a fixed redshift.

\begin{figure*}\centering
  \includegraphics [width=\linewidth] {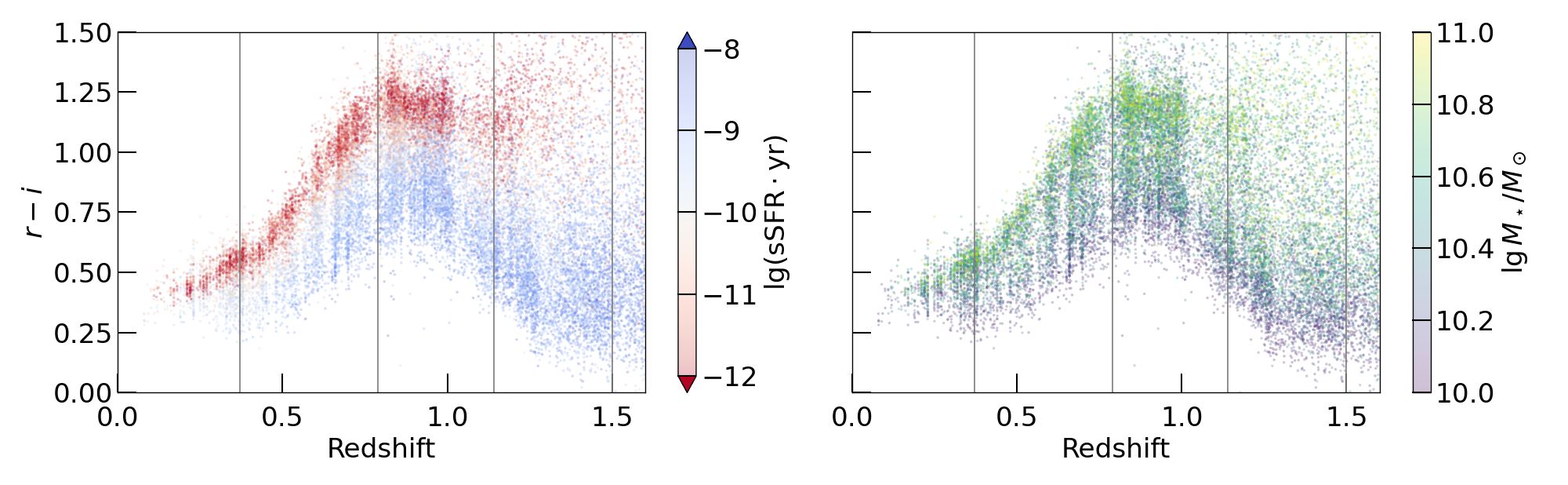}
  \vspace{-16 truept} 
  \caption[Galaxy $(r-i)(z)$ colored by sSFR and $\lg M_\star / M_\odot$]{
    Distributions of $r-i$ photometric color over redshift in the COSMOS2015 dataset, colored by derived catalog properties of specific star formation rate (left) and stellar mass (right), for a stellar mass-limited sample complete above $\lg M_\star / M_\odot = 10$. 
    As in Figure~\ref{fig:z_transitions}, vertical grey lines indicate redshifts at which the $4000~\Angstrom$ break leaves each photometric band.  
    Red Dragon's identification of RS and BC components is shown in  Figure~\ref{fig:rmi_z_Pred}. 
  }
  \label{fig:rmi_wide}
\end{figure*}

\subsection{Redshift limits} \label{sec:cuts/z}
In order to trust population characterization, we limit our focus to only those redshift bins in which reside a sufficient population of galaxies for both RS and BC. As red fractions tend to be less than half, we limit ourselves to only those redshift bins which contain a sizeable quantity of low sSFR galaxies. We use the simple definition from \citet{Ilbert+13} of 
\begin{equation} \label{eqn:qui}
  \lg {\rm sSFR \cdot yr} < -11
\end{equation}
as a definition of quiescent galaxies 
and exclude redshift bins with fewer than 75 
quiescent galaxies. In practice, this ensures a sufficient population exists within each bin for both RS \& BC. 

These requirements limit the redshift extent of each mass bin from the mass-complete limits shown in Figure~\ref{fig:Ngal_z} down to $z \in [.05, .45)$, $z \in [.1, .75)$, and $z \in [.2, 1.6)$ for the mass bins in increasing order (shown as solid points on the figure).

\subsection{Color selection} \label{sec:cuts/col} 
In an analysis of DES Y3 weak lensing data, 
\citet{Myles+21} defined color limits to exclude extremal colors (unphysical measurements assumed to be caused by catastrophic flux measurement failures). 
Mirroring their selection, we restrict each color of our primary color vector 
\begin{equation} \label{eqn:col_griz} 
  \vec c = [ g-r, \, r-i, \, i-z ]
\end{equation}
and of our extended primary color vector 
\begin{equation} \label{eqn:col_all} 
  \vec c = [ u-g, \, g-r, \, r-i, \, i-z, \, z-J, \, J-H, \, H-K_S ]
\end{equation}
to exclude galaxies without the range $[-1.5, 4.0]$. 
(These bounds turn out to be quite generous: after measuring the redshift evolution of mean colors and color scatter for each sample, we find that even five-sigma scatter off of mean colors $\vec c \pm 5 \vec \sigma$ lies entirely in the range $(-.75, 3.0)$ across all redshifts considered.) 
This cut removes $\sim 1\%$ of galaxies from our main samples. 

To exemplify a typical color distribution of galaxies, 
Figure~\ref{fig:rmi_wide} displays $r-i$ color across redshift for a mass-complete sample of galaxies, 
as colored by the physical properties of specific star formation rate (sSFR) and decimal log stellar mass ($\lg M_\star / M_\odot$). 
Vertical lines at $z = \{ .371, \, .790\}$ indicate redshift transitions of the $4000~\Angstrom$ break from $g \rightarrow r$ and $r \rightarrow i$ (see Table~\ref{tab:break_locations}). As the break drifts through $r$-band, the $r-i$ slope of quiescent galaxies with redshift rises. Once the break leaves $r$-band, $r-i$ color peaks and drifts down.

Compared to star-forming galaxies, quiescent galaxies tend to be redder across all redshifts as well as more massive; however, especially at high redshifts, some low-mass star-forming galaxies appear redder in $r-i$ than the core RS, indicating significant scatter in observed color at fixed sSFR and $\lg M_\star / M_\odot$. Therefore, single-color cuts (e.g. drawing a line in $r-i$ to divide RS from BC) are likely to confuse some low-mass and low-sSFR galaxies as RS galaxies. This issue is ameliorated by Red Dragon's multi-color GMM selection of the RS.

\section{Results} \label{sec:RDii_Results}
In this section, we discuss results of running Red Dragon (RD) on DES deep-field data in the COSMOS patch. 
In \S\ref{sec:fit} we give RD fit parameterization for the main DES bands $griz$, including component weights $w$, mean colors $\mu$, color scatters $\sigma$, and correlations $\rho$ between photometric colors for both RS and BC galaxy populations. 
In \S\ref{sec:P_red} we employ the resulting RD fit to characterize RS membership probabilities $P_{\rm RS}$; using these probabilities to define populations, we measure median sSFR and mean galactic age for RS \& BC.

\subsection{Red Dragon fit parameterization} \label{sec:fit}

In this section, we show RD fits to DES main bands $griz$ for both RS \& BC. For simplicity of visualization and discussion, we only show red fraction $f_{\rm RS}$ and mean color evolution $\langle r-i \rangle$. 
Appendix~\ref{apx:other_colors} displays parameterizations of the other mean colors, scatters, and correlations not shown here which go into the Red Dragon model. These fits are available \href{https://bitbucket.org/wkblack/rd_des/src/master/}{on Bitbucket}.\footnote{
  Hyperlinks to trained dragons: $\lg M_\star / M_\odot \in$ \href{https://bitbucket.org/wkblack/rd_des/src/master/output/bin_8to9_3K_merged_L2K_merged.h5}{$[8,9)$}, \href{https://bitbucket.org/wkblack/rd_des/src/master/output/bin_9to10_3K_merged_L2K_merged.h5}{$[9,10)$}, \& \href{https://bitbucket.org/wkblack/rd_des/src/master/output/bin_10to11_3K_merged_L2K_merged.h5}{$[10,11)$}.
} 
To show dependence of fits with stellar mass, we simultaneously visualize these fits for each of the three aforementioned decadal stellar mass bins (see Figure~\ref{fig:Ngal_z}). 

\subsubsection{Component weights}

\begin{figure}\centering
  \includegraphics [width=\linewidth] {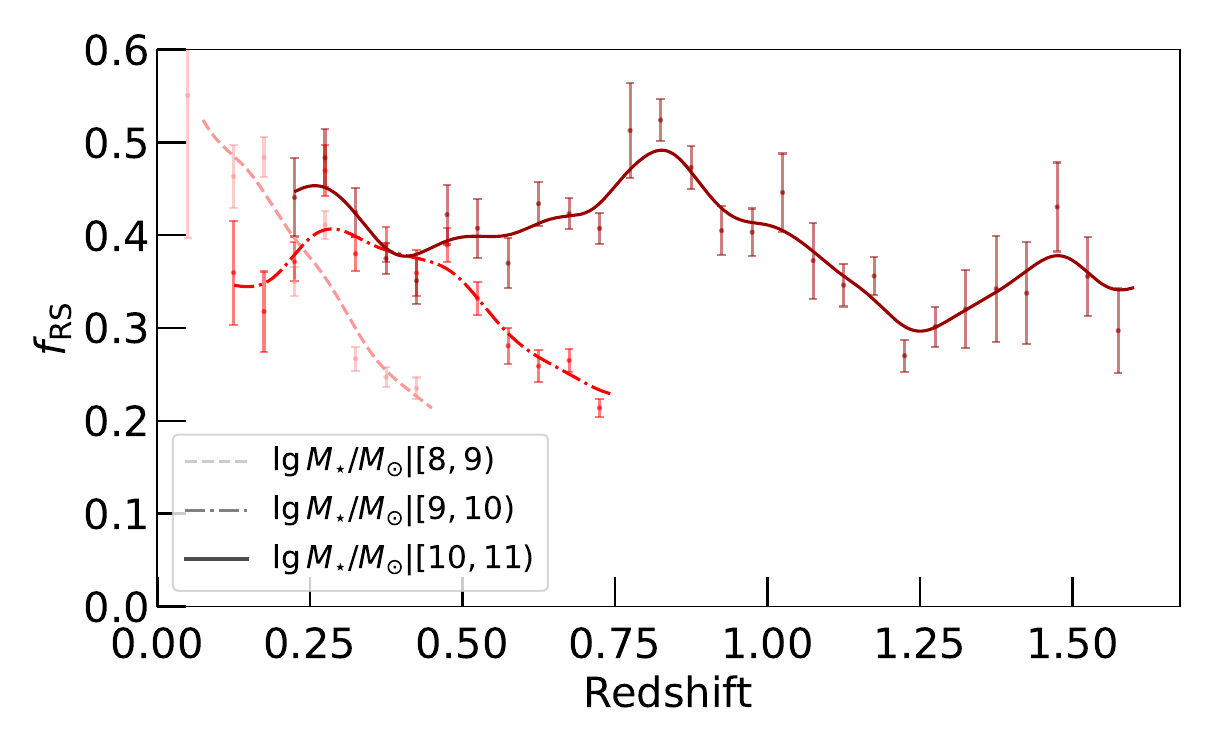}
  \vspace{-8 truept} 
  \caption[COSMOS component weights]{
    RD-measured redshift evolution of red fraction $f_{\rm RS}$ for each of the three decadal stellar mass bins (mass-complete out to the redshifts shown). 
    Each characterization uses KLLR kernel widths of $\sigma_{z} = .05$ in redshift. 
  }
  \label{fig:griz_w}
\end{figure}

Figure~\ref{fig:griz_w} shows the fraction of red galaxies $f_{\rm RS}$ as a function of redshift for each of the three decadal mass bins. For the lower two mass bins, we see clear decreasing trends, as expected due to the continual evolution of BC galaxies towards RS galaxies \citep{Butcher_Oemler_1978,Madau_1996,Connolly_1997,Madau_Dickinson_2014}. 
We tend to see higher red fractions in higher-mass bins, consistent with previous findings \citep{Baldry+04,Balogh+04,Peng+10}. 

Though in the highest-mass bin we find a net negative slope ($df_{\rm RS}/dz = -.09 \pm .03$; red fractions left of $z=.8$ are larger than those to the right), the population exhibits less consistent redshift evolution compared to the lower-mass bins. 
As high-mass galaxies concentrate towards cluster cores, red fraction at high masses fluctuates significantly due to cosmic variance. Because the COSMOS field subtends a relatively small patch on the sky, matter density fluctuates significantly within its pencil beam. 

At $z \sim .8$, the COSMOS patch spans only $\sim 30~{\rm Mpc}$ across---less than half the scale of homogeneity at that redshift \citep[$\sim 79~{\rm Mpc}$; see][]{Avila+22}. 
Near this redshift, we see number densities of BC galaxies consistent with expectations at higher redshifts, but the number density of RS galaxies abruptly rises, by about a factor of two.
This suggests a surplus of large-scale structure near $z = .8$ causes the peak in $f_{\rm RS}$ for the high mass sample. 
Indeed, the peak roughly corresponds to two X-ray bright galaxy groups near $z=.730$ and $z=.836$, which shine roughly five times brighter than other groups in the field \citep[see][Figure~9]{Leauthaud+09}. 
Cosmic variance significantly distorts red fractions in the high-mass sample as massive red galaxies almost always reside in cluster cores.

\subsubsection{Mean colors}
If we use a step function at the 4000~$\Angstrom$ break to approximate a galaxy's spectrum and use symmetric neighboring square bands, then the resulting photometric color (as a function of redshift) is a single triangle wave. The color rises from zero as the break moves into the shorter-wavelength band then falls back towards zero as the break moves through the longer-wavelength band. 
Despite the extreme simplicity of this model, it roughly explains the locations of the observed major peak of each color.

\begin{figure}\centering
  \includegraphics[width=\linewidth]{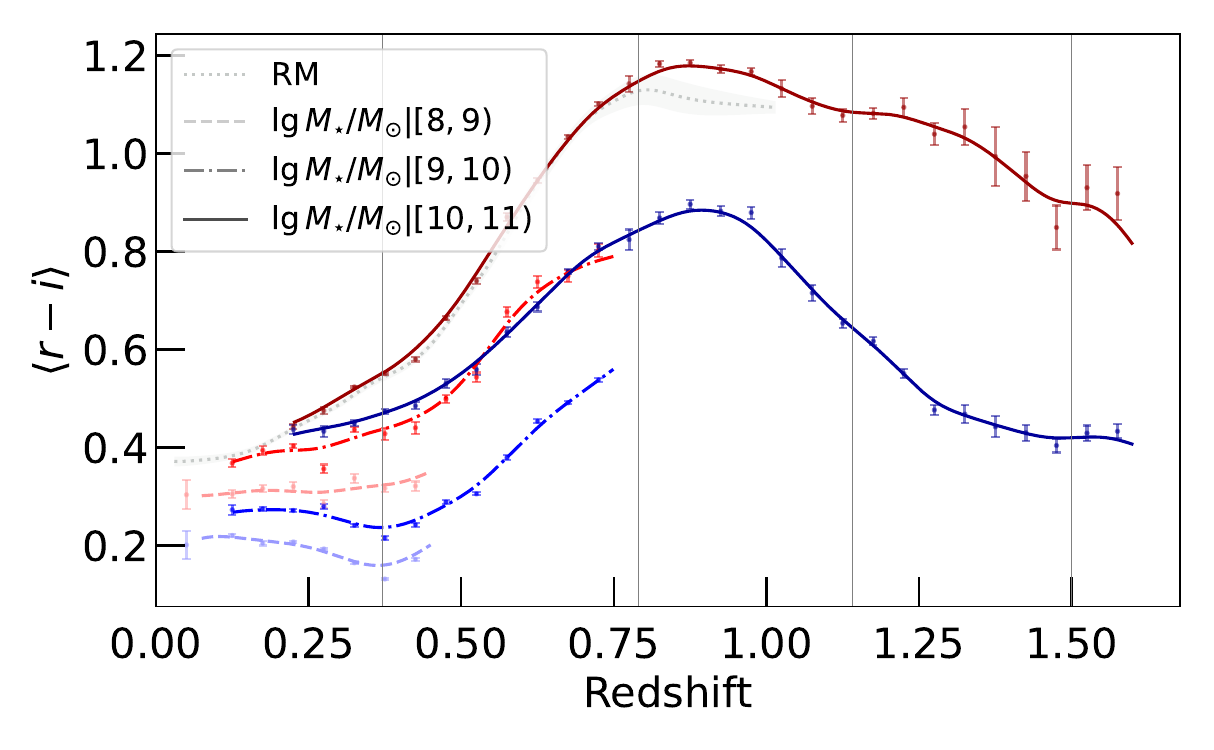}
  \vspace{-16 truept} 
  \caption[COSMOS mean color, $r-i$]{
    RD-measured mean component $r-i$ color for each decadal mass bin. RS shown in shades of red; BC shown in shades of blue; coloring grows lighter with decreasing mass. 
    The dotted silver line is a fit from the redMaPPer algorithm to the RS; as colors are magnitude-dependent, we show a spread of RM color fits based on the mean and scatter of magnitudes in our highest-mass sample. 
    As in Figure~\ref{fig:z_transitions}, vertical lines indicate $4000~\Angstrom$ exit redshifts for $griz$ bands. 
    See Figures~\ref{fig:griz_mu_gmr} \&~\ref{fig:griz_mu_imz} for $\langle g-r \rangle$ and $\langle i-z \rangle$ characterizations. 
  }
  \label{fig:griz_mu}
\end{figure}

Figure~\ref{fig:griz_mu} shows RS \& BC mean $r-i$ color evolution for each mass bin. Its behavior very roughly matches the simple step function spectral model above: color rises at $z \sim .4$ (near where the $4000~\Angstrom$ break enters $r$ band), peaks at $z \sim .8$ (near where the break exits $r$ band and enters $i$ band), and falls afterward. 

Appendix~\ref{apx:other_colors} shows fits for $\langle g-r \rangle$ and $\langle i-z \rangle$ (as well as fits for intrinsic color scatter and inter-color correlations). 
In each color, the high-mass samples closely follow Buzzard fits from \citetalias{Black+22} (up to its terminal redshift of $z=.84$), each diverging by typically $\lesssim .1~{\rm mag}$. 
We find that both RS and BC mean colors redden monotonically (albeit nonlinearly) with increasing galactic stellar mass, mirroring observed rest-frame $u-r$ behavior \citep{Baldry+04, Balogh+04}.

We compare RS mean color and scatter to DES Y3 measurements made by {\sc redMaPPer} (RM; fits made available by E.~Rykoff 2022, private comm.). 
Because RM takes into account the color--magnitude slope (whereas RD does not), we display expected mean and scatter of color based off of the corresponding mean and scatter of magnitudes
in the highest-mass sample. 
Even in our highest mass bin, 27\% of the galaxies are fainter than the RM minimum luminosity threshold \citep[one-fifth the characteristic $z$-band luminosity; see][]{Rykoff+14}; we exclusively compare to our highest mass bin (avoiding extrapolation to lower masses). 
Despite the RM sample coming from cluster members and the RD sample coming from field members, each RD-measured color traces the RM fit well for $z \lesssim .75$; beyond that redshift, $\langle g-r \rangle_{\rm RS}$ diverges towards bluer colors. 
This suggests that, at a given stellar mass, mean RS \& BC colors are independent of local density, in agreement with \citet{Balogh+04}.

\subsection{Red Dragon fit results} \label{sec:P_red}
With the photometric fits to the galaxy populations in hand, we can calculate component membership likelihoods for each galaxy. In particular, we focus on the RS membership probability $P_{\rm RS}$, 
using $P_{\rm RS} \geq .5$ to distinguish RS from BC galaxies. 
Such binary classification allows us to measure redshift evolution of sSFR values and galactic age for each population, characterizing both quiescent galaxies as well as tracing the star-forming main sequence.

\subsubsection{Galaxy selection and characterization}

\begin{figure}\centering
  \includegraphics [width=\linewidth] {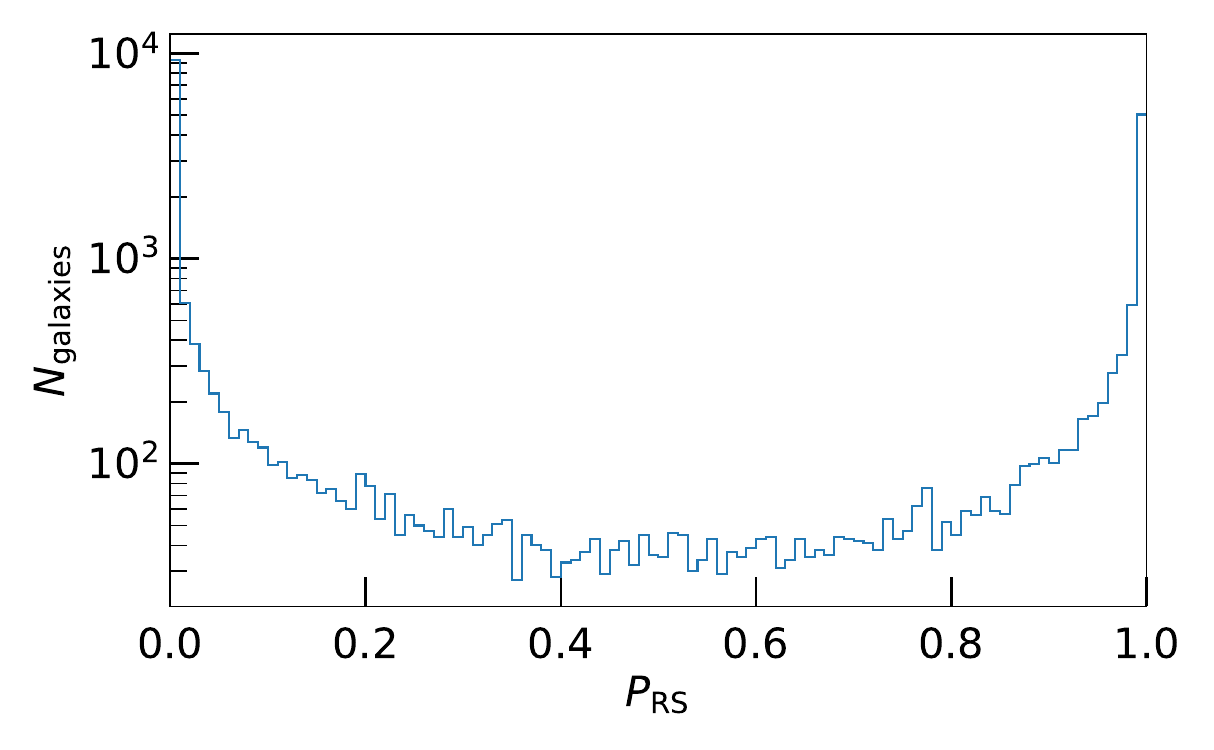}
  \vspace{-16 truept} 
  \caption[Histogram of $P_{\rm RS}$ for highest-mass dragon]{
    Histogram of the red sequence likelihood values, $P_{\rm RS}$, for decimal log mass $\lg M_\star / M_\odot \in [10,11)$. 
    A bin width of .01 highlights the order of magnitude increases at $P_{\rm RS} < .01$ and $P_{\rm RS} > .99$ , indicating high certainty of characterization for the majority of galaxies. 
  }
  \label{fig:Pred_hist}
 \vspace{5 truept} 
 \end{figure}

Using the RD parameterization above, equation~\eqref{eqn:P_alpha} ascribes to each galaxy a red sequence membership probability $P_{\rm RS}$. 
Figure~\ref{fig:Pred_hist} shows a histogram of these values for the highest mass bin, revealing strong bimodality. 
Using maximum probability $P \equiv \max(P_{\rm RS}, \, P_{\rm BC})$, we find $\sim 90\%$ of galaxies across our three mass bins have $P > 0.75$ and $\sim 75\%$ of galaxies have probabilities $P > 0.90$; this distribution lends itself naturally towards binary selection of the RS and BC. 
As maximum likelihood values $\mathcal{L}_{\max} \equiv \max(\mathcal{L}_{\rm RS}, \mathcal{L}_{\rm BC})$ decrease, galaxies are more likely ascribed middling values of $P_{\rm RS}$. For the large majority of galaxies, RD picks out two populations with distinctly different astrophysical properties.

\begin{figure}\centering
  \includegraphics [width=\linewidth] {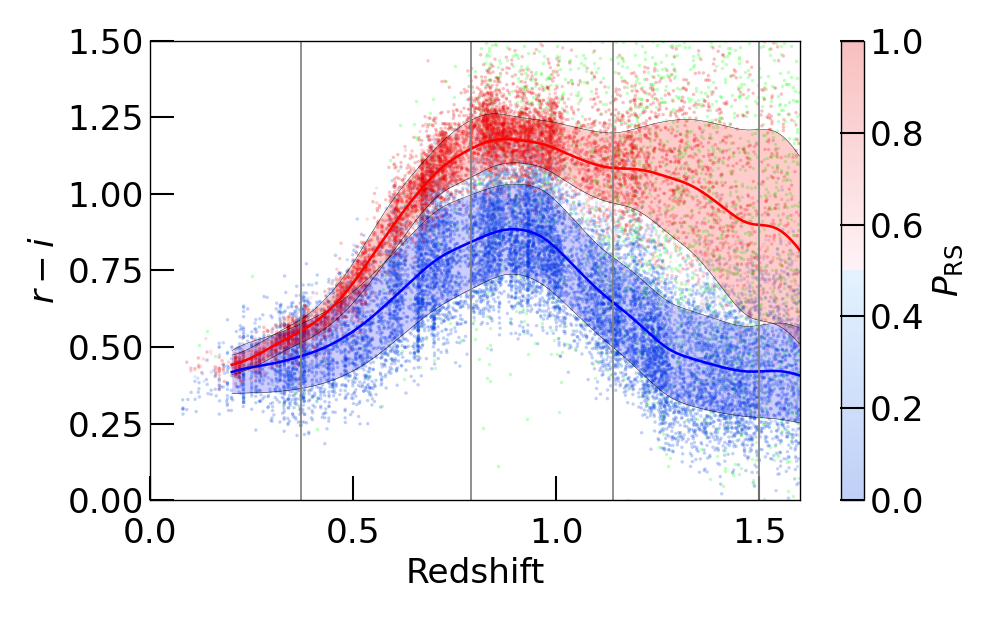}
  \vspace{-16 truept} 
  \caption[Galaxy $(r-i)(z)$ colored by $P_{\rm RS}$]{
    As Figure~\ref{fig:rmi_wide} but with points colored by RS membership probability,  $P_{\rm RS}$. 
    Green points indicate galaxies in the lowest decile of $\mathcal{L}_{\max}$ (strongly disfavored to be either RS or BC members). Red and blue contours give mean and $\pm 1 \sigma$ scatter of color for the RS and BC components, respectively, parameterized by Red Dragon for the highest stellar mass bin. 
  }
  \label{fig:rmi_z_Pred}
\end{figure}

Figure~\ref{fig:rmi_z_Pred} shows the characterization of $P_{\rm RS}$ for galaxies in the space of $r-i$ across redshift, showing the strong bimodality of probabilities. Visually compared to Figure~\ref{fig:rmi_wide}, this shows that the RD-defined RS largely captures the quiescent population. 
Green points represent galaxies which fit well to neither population; galaxies with the lowest 10\% of $\mathcal{L}$ values were neither likely candidates for RS nor BC, suggesting erroneous photometry or membership to some third component. These low-likelihood galaxies tend towards higher redshifts and had photometric color uncertainties typically $\gtrsim 2.7$ times larger than higher-likelihood galaxies. Intrinsic scatter of this low-$\mathcal{L}$ sample far exceeds that of the RS or BC, indicating that these galaxies tend to be outliers in color space.

\subsubsection{Evolution of star formation rates} \label{sec:sSFR_evolution}
Using binary selection of the RS and BC from $P_{\rm RS}$ (including low-$\mathcal{L}$ galaxies), we characterize specific star formation rates for RS and BC as functions of redshift. To avoid being skewed by extreme outliers caused by near-zero SFR values (floor: $\lg {\rm SFR \cdot yr / M_{\odot}} = -99$), we compute the median specific star formation rate, rather than the mean. 

\begin{figure*}\centering
  \includegraphics [width=.49\linewidth] {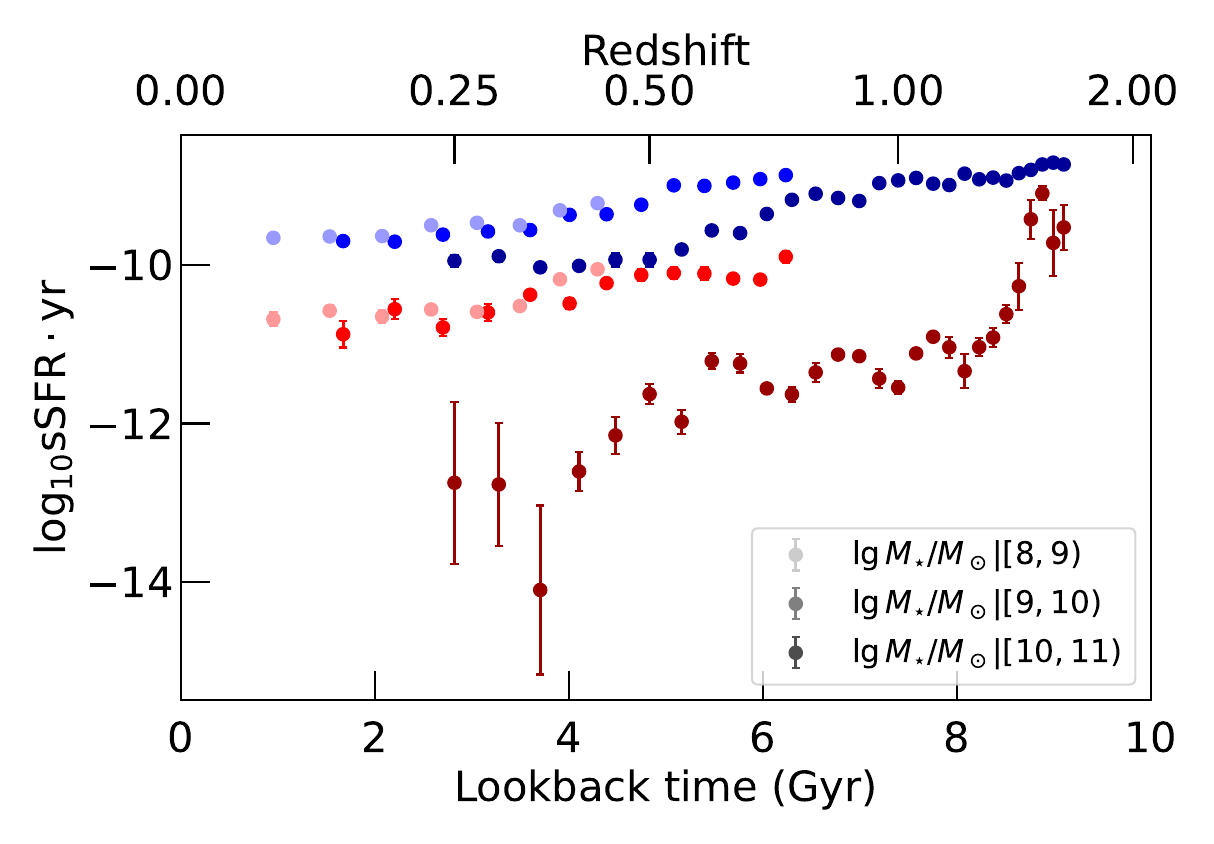}
  \includegraphics [width=.49\linewidth] {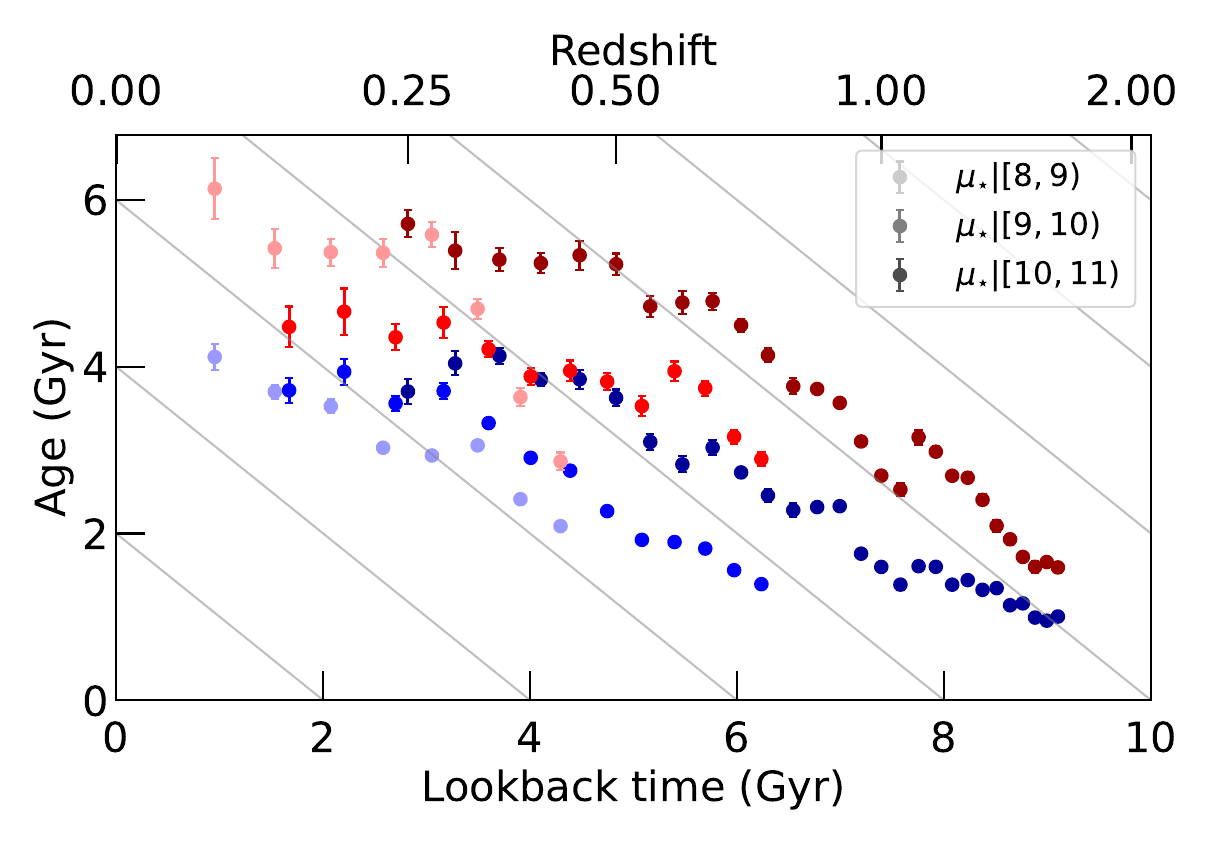}
  \vspace{-8 truept} 
  \caption[Median sSFR and mean Age]{
    {\bf Left:} 
      Evolution of the median sSFR (with bootstrap uncertainty) for RS (red points) and BC (blue) components in each of the three mass-binned samples. 
    {\bf Right:} 
      As left, but for mean galactic age. Diagonal grey lines indicate equal-epoch growth. A quenching timescale of $\mathcal{O}(1~{\rm Gyr})$ roughly marks the RS--BC separation in age. 
  }
  \label{fig:sSFR_z_mu}
\end{figure*}

Figure~\ref{fig:sSFR_z_mu}, left panel, shows median sSFR values for each population (with bootstrap uncertainties). Specific star formation rates decrease with galactic stellar mass and increase with redshift. Each mass bin shows a clear distinction in sSFR values, with quiescent galaxies a factor of 10 to 100 below star-forming galaxies (higher masses tend towards a larger separation in sSFR values). Star formation rates increase with redshift, indicating more active populations in the past (towards cosmic noon, at $z \sim 2$).

The star-forming main sequence (SFMS) roughly follows
\begin{equation} \label{eqn:SFMS}
  \lg ({\rm sSFR \cdot yr})_{\rm BC} = -10 + 
  (.2~{\rm Gyr}^{-1}) \, t 
\end{equation}
(where $t$ is lookback time). 
The time slope of this relation is in rough agreement with findings from \citet{Speagle+14}, 
who found the time slope ranged in $[.098, .176]~{\rm Gyr}^{-1}$ over our mass range. 
A more precise quiescent definition than equation~\eqref{eqn:qui} could use 1~dex below this fit of equation~\eqref{eqn:SFMS} as a dividing line between RS and BC, allowing for time evolution of the quiescent population. 
However, as discussed in \citet[][see \S7.2 therein]{Leja+22}, enforcing a hard cut selection of the RS and BC using a threshold in sSFR has considerable downsides: as sSFR is skew-lognormal (rather than bimodal, as are galactic colors), a slight shift in threshold can lead to a significant shift in population characterization. This favors using photometry to characterize population sSFR values, rather than the converse, of using sSFR values to characterize RS \& BC populations.

The uptick in median sSFR beginning at $z \sim 1.4$ (just before the $4000~\Angstrom$ break leaves $z$-band at $z=1.5$) in the high-mass population may be caused by the lack of quiescent galaxies at high redshifts. 
While this could be an actual feature (perhaps related to cosmic noon), this is similar to what would be expected due to a lack of quiescent galaxies, that the RS approaches a noise term which looks more similar to the BC. 
In the last five redshift bins displayed in Figure~\ref{fig:sSFR_z_mu}, quiescent galaxies (using the definition of eqn.~\eqref{eqn:qui}, thus differing from RD characterization) make up only $\sim 10\%$ of the population (decaying sharply with redshift beyond that point). At such a low red fraction, it becomes increasingly difficult for GMMs to detect the RS, as they favor fitting noise terms or sub-dividing the BC instead of characterizing the RS. 
Without sufficient high-quality data to form a sizeable RS population at high $z$, it will be difficult to definitively discern the uptick's cause. 

\subsubsection{Evolution of galactic age} \label{sec:age_evolution}
Similar to the previous section, we map out mean estimated galactic age for each population (typical length of time since the galaxy's formation) in the right panel of Figure~\ref{fig:sSFR_z_mu}. 
As individual galaxy ages lack uncertainty estimates, these findings should be interpreted with discretion. 
We find that RD-selected RS galaxies in each mass bin are consistently $\mathcal{O}(1~{\rm Gyr})$ older than BC galaxies, matching the expected quenching timescale \citep{Bell+04,Blanton_2006}. 

At a fixed lookback time higher-mass galaxies tend to be older than lower-mass galaxies, particularly so for the BC. Such a hierarchy is consistent with the concept of \textit{downsizing}: higher-mass galaxies tend to have older stellar populations than lower-mass galaxies \citep{Thomas+02, Nelan+05, Papovich+06}. 
The more massive a galaxy, the earlier and more rapidly it tends to have formed. 

If all galaxies in a certain stellar mass bin were born at the same time---and didn't move between RS \& BC nor grow enough to leave their mass bin---then we would expect a slope of negative one on the plot for each of the six groups, indicated by the diagonal grey lines. 
Slopes tend to be slightly shallower than the assumption of a single creation epoch and unchanging population membership. This indicates that the populations are being joined by younger galaxies (more recently formed) or that older galaxies are leaving the mass bin. (In contrast, steeper slopes indicate the reverse, that the mass bin is either joined by older galaxies or that younger galaxies are leaving the mass bin.) 

Focusing on the high-mass sample (dark points), RS and BC show distinct trends, even out to high redshift. This is somewhat in contrast to the high-mass high-redshift regime of the sSFR plot (left panel), where the RS seems to move towards the BC in what could be a degradation towards noise or a sub-division of the BC. The significant distinction of ages in the right panel indicates that Red Dragon is still selecting significantly different populations in such circumstances, even if the sSFR values diverge less than at lower redshifts.

\section{Discussion} \label{sec:RDii_Discussion} 
In this section, we 
  discuss the accuracy wherewith RD selects the quiescent population (\S\ref{sec:bACC}), 
  present rest-frame color scatter (\S\ref{sec:color_scatt}), and
  deliberate the choice of using two components versus more to characterize galaxy populations (\S\ref{sec:BIC}).

\subsection{Accuracy in selecting the quiescent population} \label{sec:bACC}

Rather than use RD selection of the RS as a `truth' to characterize median sSFR values, we can use sSFR values as a `truth' wherewith to define the RS, then measure RD's ability to select this quiescent population. 


\citet{Ilbert+13} used $\lg {\rm sSFR \cdot yr} < -11$ as a truth label for the RS and measured selection accuracy as a function of redshift. 
Their two-color hard-cut selection of the RS results in a \emph{balanced accuracy} (bACC; the average of specificity and sensitivity) consistent with linearly decay over redshift, moving from $\sim 95\%$ accuracy at redshift zero down to $\sim 65\%$ accuracy at $z=2.5$ (random selection results in a bACC of 50\%). Their sample extends to redshift $z=3$. 

To match their redshift extent in our sample requires that we leave mass completeness and fit redshift bins with incredibly sparse populations of quiescent galaxies. To increase RS size, we fit all $\lg M_\star / M_\odot > 9$ galaxies here, rather than limit ourselves to a single mass decade. Because this widening of mass range permits significant GMM parameter drift, we perform a three-component fit to account for stellar mass dependence of population parameterizations. 
Though these fits may degrade somewhat towards higher redshifts (as discussed in \S\ref{sec:sSFR_evolution}), we will show that the fits retain utility nonetheless.

\begin{figure}\centering
  \includegraphics [width=\linewidth] {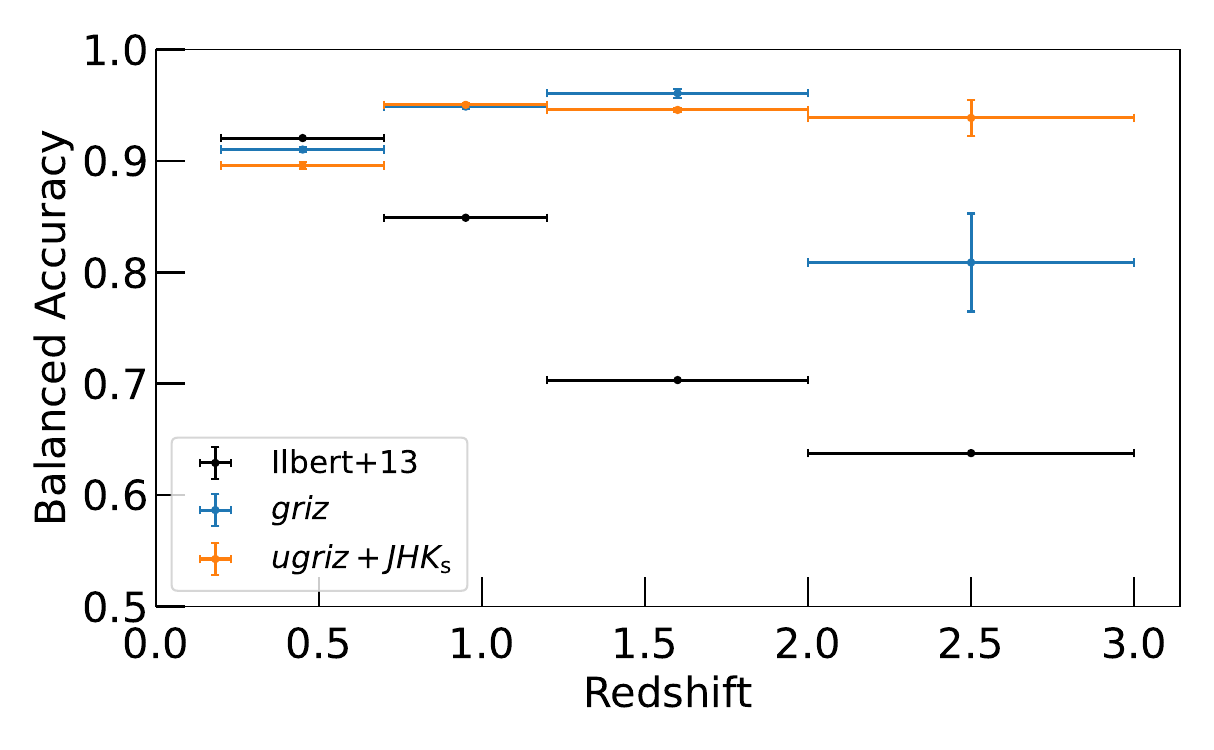}
  \vspace{-16 truept} 
  \caption[bACC of selecting quiescent galaxies]{
    Balanced accuracy in selecting quiescent galaxies (see equation~\eqref{eqn:qui}). 
    \citet{Ilbert+13} selection shown in black (no uncertainties available). 
    RD selection shown in blue and orange for DES main bands ($griz$) and extended photometry ($ugriz+JHK_{\rm s}$), respectively, with three-component dragons applied to a $\lg M_\star / M_\odot > 9$ sample. 
  }
  \label{fig:bACC_Ilbert+13}
\end{figure}

\begin{figure*}\centering
  \includegraphics [width=.49\linewidth] {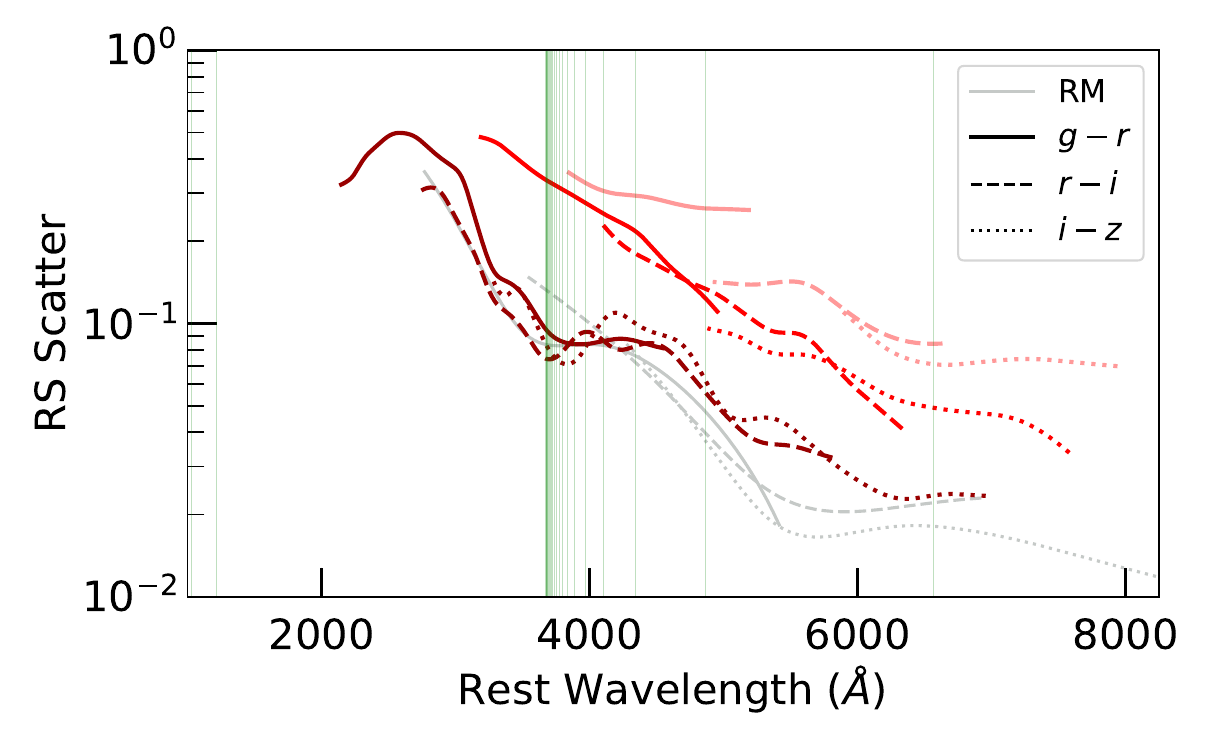} 
  \includegraphics [width=.49\linewidth] {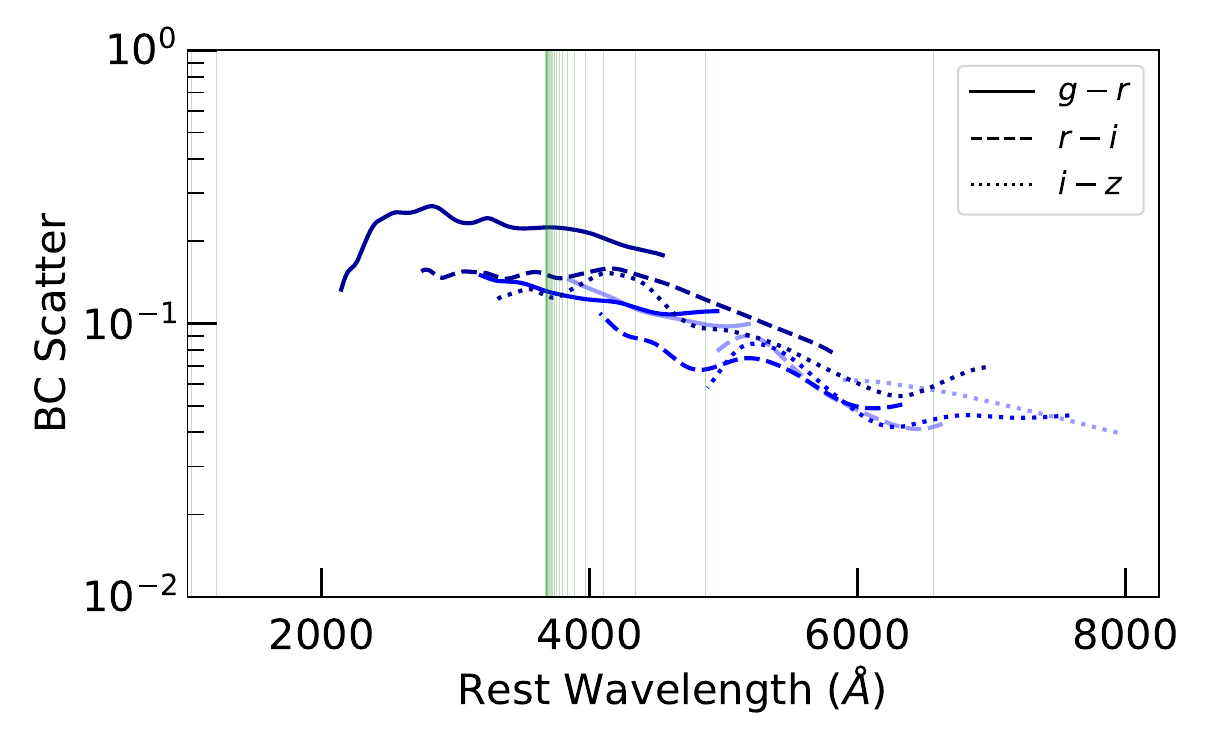} 
  \vspace{-8 truept} 
  \caption[Colors scatters vs imputed rest-frame wavelength]{
    RS (left) \& BC (right) color scatter as functions of imputed rest-frame wavelength $\lambda_{\rm rest} = \lambda_{\rm tr, col} / (1+z)$. 
    Lighter colors indicate lower-mass bins, as in Figure~\ref{fig:griz_mu}, while line styles vary for different photometric colors. 
    Silver lines indicate RM fit results. 
    Green solid lines show hydrogen spectral series lines, with most belonging to the Balmer series (beginning with H$\alpha$ at $6550~\Angstrom$ and terminating at $3645~\Angstrom$). 
    Note that at a fixed imputed rest wavelength, each color scatter is measured at a different redshift, with $g-r$ at the lowest redshift and $i-z$ at the highest (redshift increases towards shorter rest wavelengths).
  }
  \label{fig:scat_rest}
\end{figure*}

Figure~\ref{fig:bACC_Ilbert+13} compares balanced accuracy of two-color hard-cut selection \citep[][black errorbar points; no uncertainties in bACC available]{Ilbert+13} to Red Dragon's multi-color selection using DES main photometry $griz$ (blue points) and additionally extended photometry $ugriz + JHK_{\rm s}$ (orange points) models of the RS. 
In contrast to the linearly decaying RS selection accuracy of \citet{Ilbert+13}, Red Dragon consistently selects the RS with high accuracy. Across the board, we find ${\rm bACC} \sim 95\%$ for the two dragons, with the extended-photometry dragon outperforming the $griz$ dragon significantly at the highest redshifts. 
Using RD in the intermediate redshift bins spanning $z\in(0.7, 2.0)$ yields significant gains in accuracy compared to the two-color model. Particularly at $z \sim 1.6$, we see a gain in accuracy of $\sim 25\%$. 
Though the highest redshift bin has very few quiescent galaxies, and therefore has wider uncertainty in bACC, we still see significant gains in accuracy. 
We therefore see clear superiority in GMM selection of the RS as compared to using hard cuts in color--color space, especially at higher redshifts.

The skew-lognormal distribution of sSFR gives hard cut selection of the RS vs BC considerable dependence on the particular sSFR threshold used. 
This is particularly visible in the highest redshift bin. Using equation~\eqref{eqn:qui}, only 58 of 35,216 galaxies are quiescent (roughly 1 in 600 galaxies). However, using $1~{\rm dex}$ below equation~\eqref{eqn:SFMS} as a quiescent definition (merely adding time dependence), the count drastically increases, to 12,422 (roughly 35\% of galaxies). 
Even varying the truth threshold by only $.3~{\rm dex}$ (a factor of two) results in a $\sim 5\%$ change in bACC values---a small yet significant shift. This sensitivity to sSFR threshold makes discussing {\it relative} bACC values more productive than discussing {\it absolute} bACC values. 
Rather than focus on a ``quiescent accuracy'' of RS selection, we find more utility in using the RS to measure quiescence of its constituent members, as was performed in \S\ref{sec:sSFR_evolution}.

\subsection{Color scatter in rest frame} \label{sec:color_scatt}
While photometric colors are measured over large swaths of a galaxy's observed spectrum, a color measured at a particular redshift could be interpreted as a smoothed spectral slope, taken about a certain rest-frame wavelength. Color scatters then measure variation in the smoothed slope at that same rest wavelength. 

This would be an exact interpretation if photometric filters were infinitesimally narrow box functions; photometric color would then precisely measure an instantaneous log spectrum slope at the transition wavelength between the bands---the point where one filter begins capturing more light than another. 
A measurement of a given color or scatter at redshift $z$ then corresponds to a rest wavelength feature at 
$\lambda_{\rm rest} = \lambda_{\rm tr, col} / (1 + z)$, 
where $\lambda_{\rm tr, col}$ is the transition wavelength for a given color
(listed in Appendix~\ref{apx:Z_transition}). 
For our photometry, each filter has substantial width as well as unique asymmetries, so colors and scatters are smoothed and somewhat distorted from a measurement made with infinitesimally narrow box-function filters. 
Despite these imperfections, interpreting colors and scatters at imputed rest-frame wavelength shows good agreement across redshift and filter choice.


Figure~\ref{fig:scat_rest} shows color scatters transformed from their measured domain of redshift (as shown in Figures~\ref{fig:griz_sig}, \ref{fig:griz_sig_gmr}, \& \ref{fig:griz_sig_imz}) to an imputed rest-frame wavelength $\lambda_{\rm rest}$. 
%
We see for both RS \& BC that scatter tends to decrease towards longer wavelengths in the range surveyed here. 
Scatter in the RS monotonically decreases with increasing galactic stellar mass. In contrast, the BC shows no significant trend with mass, with its scatter consistent to roughly a factor of two across all wavelengths surveyed here. 
While \citet{Baldry+04} showed that these trends hold for rest frame $u-r$, we show here the trend holds true more generally, across the entire wavelength domain surveyed here.

While it is beyond the scope of this paper to definitively prove the causes of the scatter, the results are largely in agreement with expectations. 
In the highest-mass bin of the RS, we see a significant leap in scatter at wavelengths $\lambda < 4000~\Angstrom$, likely caused by differences in metallicity, age, and sSFR, which can cause drastic differences in ultraviolet slope \citep[see e.g.][]{Kriek+11}. 
While we expect dust only marginally affects the RS scatter, we expect it to substantially impact the BC scatter. 
In our wavelength ranges, variation in dust content of BC galaxies effects BC color scatter more at shorter wavelengths, where smaller particles of dust are more capable of scattering light \citep{Fioc+19}. 
Preliminary results from SPS modeling shows that at wavelengths shorter than $8000~\Angstrom$, color scatter caused by dust typically exceeds any color scatter caused by differences in metallicity. 
We leave more detailed analysis of color scatter at a given rest wavelength for future papers.


\subsection{Optimal component count} \label{sec:BIC}
Red Dragon allows for fitting of not only RS and BC, but additionally of any number of Gaussian mixture model components. Additional components model either ``green valley'' galaxies, a noisy background (e.g. from galaxies with bad photo-$z$ estimates), or even further components (e.g. dividing RS or BC into sub-populations in order to model non-Gaussianities). 

\citetalias{Black+22} quantified optimal component count using the Bayesian Information Criterion (BIC). BIC measures relative information loss of different models (run on the same data); it increases with the number of model parameters and decreases with increased maximum model likelihood. BIC thus increases with model complexity and decreases with improved fit, so models with lower BIC better minimize information loss. 
Log relative likelihoods $\ln \mathcal{L} = ({\rm BIC}_a - {\rm BIC}_b) / 2$ of model $b$ minimizing information loss as compared to model $a$ tend to have values in the hundreds, leading to probabilities of model superiority tending strongly towards zero and one (i.e. $\epsilon$ and $1-\epsilon$, with $\epsilon \lesssim 10^{-20}$). 
However, bootstrap uncertainties on BIC values tend to overshadow differences in BIC between models, such that comparison of \emph{significance} of a non-zero difference in BIC tends to be more useful than the probability itself. 
In the spirit of BIC---minimizing both model complexity as well as information loss---we only prefer a more complicated model if it \emph{significantly} diminishes information loss (i.e. if the log relative likelihood $\ln \mathcal{L}$ significantly differs from null or equivalently if BIC values differ significantly).

For neither $griz$ nor $ugriz + JHK_{\rm s}$ photometries do we find significant differences in BIC for any of our mass-binned dragons on moving from a two component model ($K=2$) to more components ($K \geq 3$). Differences in BIC were typically\footnote{
  The strongest preference for a $K=3$ model came from the highest mass bin, using the extended photometry, around $z=.75$, but this preference was only $\sim 2\sigma$ significant. Other redshifts were consistent with null preference. 
} $<1\sigma$ from zero, never crossing $3 \sigma$ significance for any sample. This lack of significance implies that the simpler, two-component model should be preferred. 
While these results hold for our mass resolution of 1~dex and our redshift resolution of $\Delta z = .05$, a $K=2$ GMM run using wider mass or $z$ bins may fail to model the populations well (as is discussed in Appendix~\ref{sec:permissible_width}). In such circumstances, a third component may be significantly favored for inclusion (as in \S\ref{sec:bACC}, where we cover over two decades of mass).

\section{Conclusions} \label{sec:RDii_Conclusion}
We employ Red Dragon, a Gaussian mixture classifier, to identify red and blue galaxy populations using multicolor photometry and derived properties from the COSMOS2015 galaxy catalog.  Components are independently fit in three decade-wide mass bins spanning stellar masses, $\lg M_\star / M_\odot \in [8,11)$, as a function of redshift using mass-complete samples.  
RD outputs mean colors and color covariance as a function of redshift for each galactic population component. A galaxy's location in color space forms the basis of component classification.  

We find that population red fractions decline with increasing redshift For the lowest two stellar mass bins, but the trend for $\lg M_\star / M_\odot > 10$ galaxies is considerably weaker.  Mean colors that include the $4000~\Angstrom$ break show characteristic non-monotonic behavior, first increasing then decreasing as the feature passes through adjacent  passbands.  We perform the first measurement of intrinsic color covariance in each sub-population, finding mostly positive values that may simply reflect variations in overall spectral slope.  


Median specific star formation rates and mean galactic ages differ between the two components.  
Within each mass bin the RD-selected RS is consistently older (by $\gtrsim 1~{\rm Gyr}$) and more quiescent (by $\gtrsim 1~{\rm dex}$) than the BC across all redshift. 
Galactic ages follow a downsizing trend with mass, such that heavier galaxies tend to be older. 
The tendency towards a shallow slope of galactic age over time indicates galaxy growth: young lightweight galaxies immigrate to a mass bin while old heavyset galaxies emigrate from that bin. 
Though using a hard cut in sSFR as a truth label for the RS has complications (as discussed at the end of \S\ref{sec:bACC}), we find high selection accuracy, which for $z>.7$ vastly outperforms typical two-color selection of the quiescent population. 

We briefly discuss scatter in color as a function of imputed rest-frame wavelength (\S\ref{sec:color_scatt}). 
This roughly gives the variation in spectral slope at a given wavelength, showing which spectral regions are more or less scattered for RS and BC. 
We find that color scatter tends to decrease towards longer wavelengths---more so for the RS than for the BC. The RS displayed mass dependence, with lower-mass galaxies exhibiting larger scatters than higher-mass galaxies across all wavelengths surveyed here. The BC showed no significant trend in mass dependence of color scatter. We leave to future papers a more detailed analysis of the causes of color scatter at various rest-frame wavelengths. 

We find no significant evidence for using more than two components to model the galaxy populations in our decadal mass bins (\S\ref{sec:BIC}). 
However, in the low-$z$ SDSS dataset used in \citetalias{Black+22}, we had found significant evidence towards using three components instead of only two. The third component consistently had lower weight ($w<10\%$), higher scatter (roughly twice that of the BC), and usually had the lowest (small positive or consistent with null) inter-color correlations. This indicates that the third component merely captures `noise', i.e. galaxies that didn't fit well in either component. 
Whether such galaxies are excluded due to low likelihoods (green points of Figure~\ref{fig:rmi_z_Pred}) or chosen as a third component of the mixture model, the two core populations of RS and BC remain dominant, with insufficient non-Gaussianities to warrant their sub-division in our samples. 

Finally, we turn to the future of the Red Dragon algorithm. In its current state, it is designed to run GMM parameters $\boldsymbol{\theta}$ with a single variable smoothly (in this paper, we run with redshift). Any other fields must be binned (in this paper, the three mass decades). 
An improved version of the algorithm could allow $\boldsymbol{\theta}$ to evolve with $N$ fields (e.g. redshift, stellar mass, and local density). This higher dimensional parameterization of the RS and BC would yield valuable insights. 
For example, it would reveal whether RS \& BC mean colors evolve with local density \citep[as observed by][]{Balogh+04} purely due to the correlation of stellar mass with local density (i.e. mean colors are invariant to local density) or whether RS mean colors do indeed depend on local density at fixed stellar mass. This could reveal a global red fraction function $f_{\rm RS}(z, M_\star, \delta)$, measuring mean red fraction as a function of redshift, stellar mass, and local density.

As telescopes including Euclid \citep{Laureijs2011Euclid}, the James Webb Space Telescope \citep{Gardner+06JWST}, 
  and future telescopes like the Legacy Survey of Space and Time \citep{Ivezic+19LSST} and the Roman Space Telescope \citep{Spergel2015WFIRST} 
increasingly yield quality high-redshift data of the quiescent population, Red Dragon will yield increasingly precise characterization of galaxy populations, leading to improved understanding of galaxy formation in our universe.

\section*{Acknowledgements}
The authors thank the anonymous reviewer for their generous suggestions in improving the paper. 

WKB thanks Peter Melchior for providing {\sc pyGMMis}, the backbone of Red Dragon, along with crucial interpretation of parameter fitting. He also thanks Johnny Esteves for editing assistance and the stellar name of ``Red Dragon'' and Eric Bell for his invaluable astrophysics insights. 
Without the support of WKB's wife Eden and son Fletcher, this work could not have been completed.

Funding from NASA Grant 80NSSC22K0476 provided crucial support for this research. 
The authors additionally thank the National Energy Research Scientific Computing Center (NERSC) for access to computing resources (including COSMOS data) used to carry out the analyses of this paper. 
Part of this work was performed at the Aspen Center for Physics, which is supported by National Science Foundation grant PHY-2210452.

This work was made possible by the generous open-source software of 
  \href{https://matplotlib.org/} {\sc Matplotlib} \citep{pyplot}, 
  \href{https://numpy.org/doc/stable/} {\sc NumPy} \citep{numpy}, 
  \href{https://docs.h5py.org/en/stable/} {h5py} \citep{h5py}, 
  \href{https://github.com/pmelchior/pygmmis} {py\-GMMis} \citep{Melchior_Goulding_2018}, and
  \href{https://github.com/afarahi/kllr/tree/master/kllr} {KLLR} \citep{Farahi+18,Farahi+22}.

\section*{Data Availability}
Algorithms used in this paper are publicly available in the \href{https://bitbucket.org/wkblack/workspace/projects/DRAG}{\texttt{DragonHoard}} project on Bitbucket.\footnote{\href{https://bitbucket.org/wkblack/workspace/projects/DRAG}{\texttt{bitbucket.org/wkblack/workspace/projects/DRAG}}} The Red Dragon algorithm resides in the repository \href{https://bitbucket.org/wkblack/red-dragon-gamma} {\texttt{red-dragon-gamma}} while other routines and results used in this paper are found in the repository \href{https://bitbucket.org/wkblack/rd_des/}{\texttt{rd\_des}}.




\href{https://cdcvs.fnal.gov/redmine/projects/des-sci-release/wiki/Y3_deep_fields_cat}{The COSMOS dataset} used in this analysis is available publicly, on the \href{https://des.ncsa.illinois.edu/releases/y3a2/Y3deepfields}{DES Data Management} page \citep{Hartley+22}. 
Zenodo hosts the vetted datasets in a RD-readable format \citep{Black23_COSMOS_data}.

\bibliographystyle{mnras}
\bibliography{main} 

\appendix

\twocolumngrid

\section{Band Characterization} \label{apx:Z_transition}

\begin{table}\centering
  \caption[Entry and exit redshifts of the $4000~\Angstrom$ break]{
    Characterization of \href{https://noirlab.edu/science/documents/scidoc0472}{DECam} ($ugriz$) and \href{https://www.eso.org/sci/facilities/paranal/instruments/vircam/inst/Filters_QE_Atm_curves.tar.gz}{VIRCAM} ($JHK_{\rm s}$) bands used in this analysis, including central wavelengths $\lambda_{\rm c}$, widths $\Delta \lambda$ (FWHM), and redshifts at which a 4000~\AA\ rest-frame wavelength source will enter ($z_{\rm 4k, EN}$) and exit ($z_{\rm 4k, EX}$) each band. 
    Note that the left edges of $u$- and $g$-bands are shorter than $4000~\Angstrom$, so their entry redshifts are negative. 
  }
  \begin{tabular}{crrcc}
    \hline 
    band & $\lambda_{\rm c}$ (\AA) & $\Delta \lambda$ (\AA) & $z_{\rm 4k, EN}$ & $z_{\rm 4k, EX}$ \\ 
    \hline 
    $u$   &   3552 &  885 & -.22 &    0 \\
    $g$   &   4730 & 1503 & -.01 & .371 \\
    $r$   &   6415 & 1487 & .418 & .790 \\
    $i$   &   7835 & 1470 & .776 & 1.14 \\
    $z$   &   9260 & 1520 & 1.12 & 1.50 \\
    \hline 
    $J$   & 12 523 & 1725 & 1.92 & 2.35 \\
    $H$   & 16 451 & 2915 & 2.75 & 3.48 \\
    $\;K_{\rm s}$ & 21 467 & 3090 & 3.98 & 4.75 \\
    \hline 
  \end{tabular}
  \label{tab:break_locations}
\end{table}

Table~\ref{tab:break_locations} characterizes the photometric bandpass filters used in this analysis, including central wavelengths $\lambda_c$, filter widths $\Delta \lambda$ (FWHM), as well as entry/exit wavelengths for the 4000~\AA\ break ($z_{\rm 4k, EN/EX}$). 
This simplified presentation of central wavelengths and widths belies the asymmetries and rounded edges of each filter's transmission, so crossing redshifts are only approximate. 
The exit and entry wavelengths don't perfectly coincide between bands, since filters may overlap (e.g. in the case of $r \rightarrow i$) or have significant gaps between them (e.g. with the VIRCAM filters). 

These entry and exit redshifts can be shifted for any other rest-frame wavelength $\lambda$ (rather than 4000~\AA): 
\begin{equation}
  z_\lambda = (z_{\rm 4k} + 1) \frac{4000~\text{\AA}}{\lambda} - 1 
\end{equation}
For example, the Balmer break at $\lambda_{\rm B} = 3647.05$~\AA\ would have shifted transition redshifts, such that $z_{\rm B} \approx 1.097 (z_{\rm 4k} + 1) - 1 = 1.097 z_{\rm 4k} + .097$ and the Lyman limit at $\lambda_{\rm L} = 911.763$~\AA\ would have $z_{\rm L} \approx 4.4 (z_{\rm 4k} + 1) - 1$. Thus Balmer transition redshifts are roughly 10\% larger than 4000~\AA\ break redshifts. 

Transition wavelengths between filters are taken as the geometric mean between bandpass edges, i.e.: 
\begin{equation}
  \lambda_{\rm tr} \equiv \sqrt{\left( \lambda_{{\rm c}, a} + \frac12 \Delta\lambda_a \right) \left( \lambda_{{\rm c}, b} - \frac12 \Delta\lambda_b \right)}
\end{equation}
for bands $a$ and $b$.\footnote{
  As filter shapes are invariant over redshift in the space of $\log\lambda$ but constrict in the space of $\lambda$, we favor the geomtric mean over the arithmetic mean. In practice, this makes $<.1\%$ difference in $\lambda_{\rm tr}$. 
} 
This yields the transition wavelengths listed in Table~\ref{tab:lambda_transition}. 
\begin{table}\centering
  \caption[Transition wavelengths]{
    Wavelengths at which a monochromatic signal moves from being picked up by one band more than another. 
    Photometric filters as in Table~\ref{tab:break_locations}. 
  }
  \begin{tabular}{c|c}
    \hline 
    bands & $\lambda_{\rm tr}$ (\AA) \\
    \hline 
    $u \rightarrow g$ & 3987 \\
    $g \rightarrow r$ & 5576 \\
    $r \rightarrow i$ & 7131 \\
    $i \rightarrow z$ & 8534 \\
    $z \rightarrow J$ & 10,800 \\
    $J \rightarrow H$ & 14,200 \\
    $H \rightarrow K_{\rm s}$ & 18,900 \\
    \hline 
  \end{tabular}
  \label{tab:lambda_transition}
\end{table}
These $\lambda_{\rm tr}$ values can then be used to impute a rest-frame wavelength for a given color, as is done in \S\ref{sec:color_scatt} and Appendix~\ref{sec:color_interp}.

\section{Remaining fits} 
\label{apx:other_colors}

Here we display the parameterizations of the main $griz$ dragons not presented in the main text (\S\ref{sec:fit}), including the remaining mean colors, scatters, and correlations. 

\subsection{Mean colors (continued)}

\begin{figure}\centering
  \includegraphics[width=\linewidth]{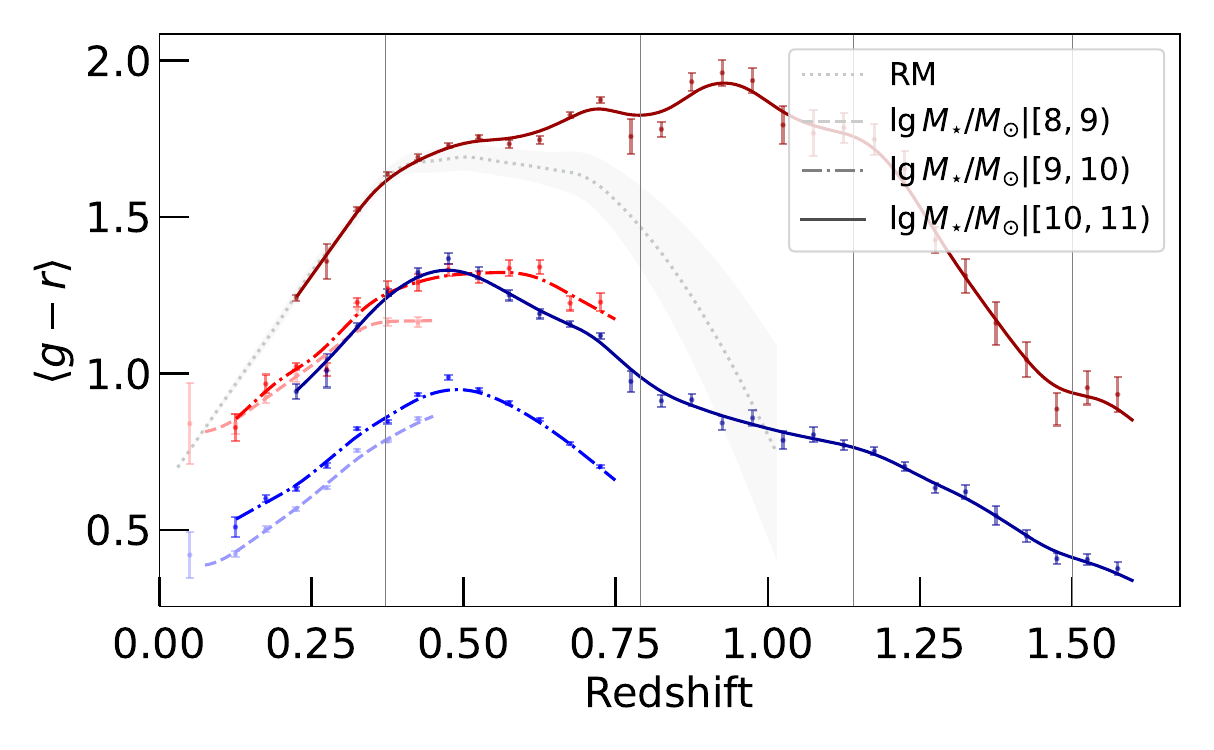}
  \vspace{-16 truept} 
  \caption{
    As Figure~\ref{fig:griz_mu}, but for $g-r$. 
  }
  \label{fig:griz_mu_gmr}
\end{figure}

\begin{figure}\centering
  \includegraphics[width=\linewidth]{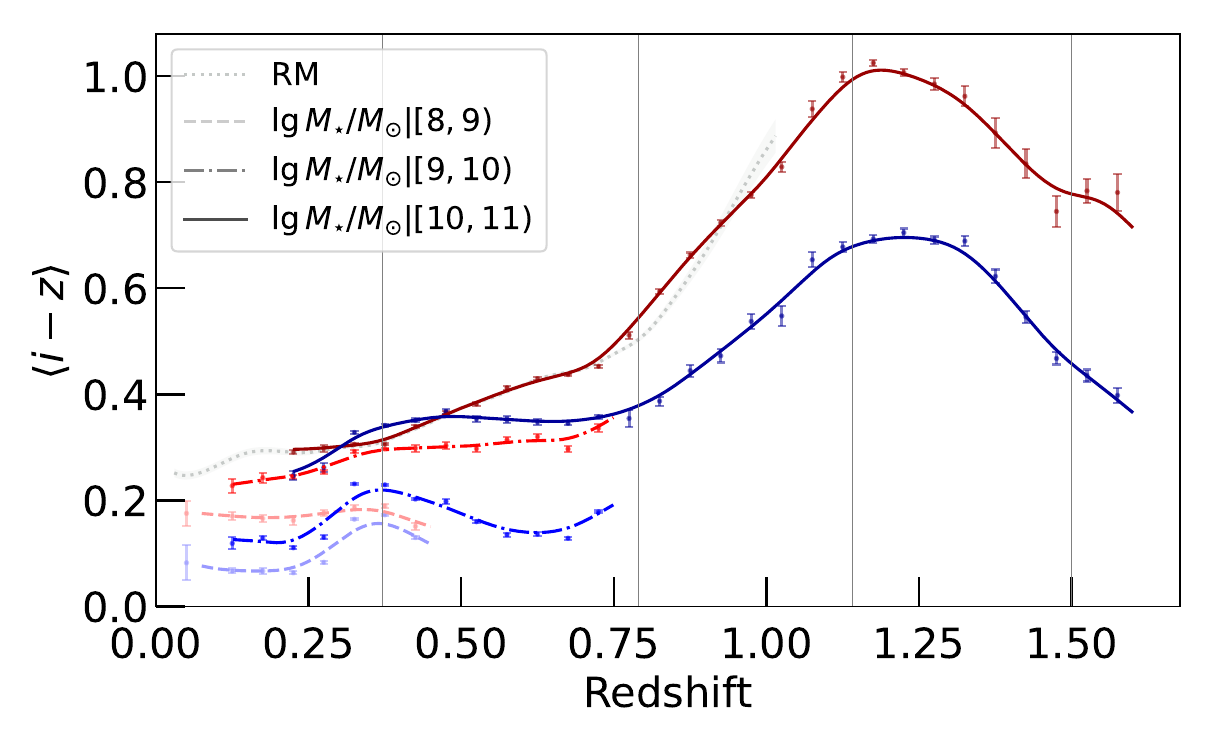}
  \vspace{-16 truept} 
  \caption{
    As Figure~\ref{fig:griz_mu}, but for $i-z$. 
  }
  \label{fig:griz_mu_imz}
\end{figure}

Figure~\ref{fig:griz_mu_gmr} shows the RD fit to the RS in $g-r$ has a $\sim 0.5~{\rm mag}$ divergence from RM near $z=1$. However, the RM-measured slope of mean color with respect to luminosity is significantly larger than this offset. Therefore, relative to Figures~\ref{fig:griz_mu} \&~\ref{fig:griz_mu_imz}, and taking into account divergence from fit relative to slope offset, the fit is still in fair agreement. 
This separation is also roughly the size of the measured RS scatter at that redshift (see Figure~\ref{fig:griz_sig_gmr}), so to $2\sigma$, the lines are in fine agreement.

\subsection{Intrinsic color scatter}

In this section, we present the plots of intrinsic color scatter in RS and BC for our model. 
To some extent, the scatter in photometric color depends on the diversity of models used for photo-$z$ estimation, so a lack of diversity of models could synthetically alter scatter. However, a significant portion of the scatter will arise from variations in metallicity and SFH, so errors in photo-$z$ estimation in this highly observed patch of the sky are unlikely to significantly corrupt color scatter estimates.

\begin{figure}\centering
  \includegraphics[width=\linewidth]{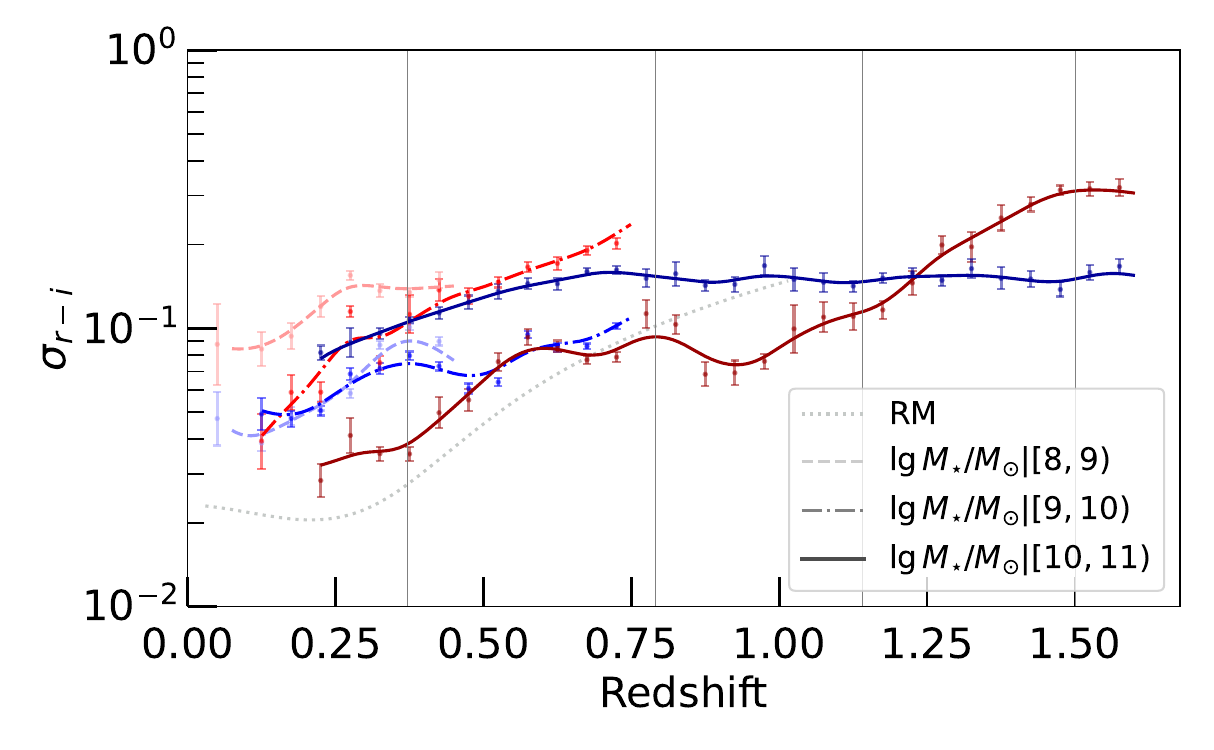}
  \vspace{-16 truept} 
  \caption[COSMOS color scatter, $r-i$]{
    RD-measured intrinsic color scatter in $r-i$; lines as in Figure~\ref{fig:griz_mu}. 
    The dotted silver line is an estimate from the redMaPPer algorithm which, unlike mean color, does not vary with stellar mass. Because RD does not include a mean color gradient with magnitude, as redMaPPer does, the scatter in the highest mass RD bin is somewhat larger than the values measured by redMaPPer. 
    See Figures~\ref{fig:griz_sig_gmr} \&~\ref{fig:griz_sig_imz} for $\sigma_{g-r}$ and $\sigma_{i-z}$ characterizations. 
  }
  \label{fig:griz_sig}
\end{figure}

Figure~\ref{fig:griz_sig} details the RD fitting of intrinsic $r-i$ scatter as it evolves with redshift for each stellar mass bin, for both RS and BC components. 
RS scatter in the heaviest stellar mass bin roughly matches the RM fit, consistent within a factor of two. This agreement between RD and RM is non-trivial: while RM trained on a seeded spectroscopic set, RD trained on a small photometric patch. 
The two lower-mass bins showed $\gtrsim 2$ times the scatter of high-mass galaxies. 
As scatter in metallicity and age increases towards lower galactic stellar masses \citep{Mannucci+10}, the scatter in color for lower-mass RS galaxies will therefore also increase, as seen here. 
In contrast, scatter in the BC shows less variation with mass, with no clear trend apparent. 

Section~\S\ref{sec:color_scatt} shows scatter of each color in imputed rest-frame wavelength. 
Across the board, scatter in RS color tends to increase with redshift (or equivalently, towards shorter rest-frame wavelengths). BC scatters tend to increase until the color's transition redshift, after which it is roughly constant. 
For the RS, scatter almost always reduces with increasing galactic stellar mass $M_\star$, but BC scatter exhibits non-monotonic behavior with mass. 
Our results broadly agree with the finding from \citet{Baldry+04}, which measured non-monotonic behavior of $\sigma_{\rm BC}$ with regard to mass for rest-frame $u-r$. 

For each color of the high-mass galaxy sample, the RS scatter begins to exceed BC scatter near the redshift where the $4000~\Angstrom$ break exits a color's longer-wavelength band (for $r-i$ this occurs at $z \sim 1.2$). 
In contrast, for the lightest galaxies, we found consistently that $\sigma_{\rm RS} \gtrsim \sigma_{\rm BC}$ for their full redshift extent. This could be caused by the plurality of quenching mechanisms present at low galactic mass; while high-mass galaxies ($\lg M_\star / M_\odot \gtrsim 10.5$) are essentially all quenched by AGN, lower-mass galaxies are quenched by a variety of mechanisms, such as supernovae, stellar winds, or reionization \citep{Wechsler_Tinker_2018}. The increased variety in quenching mechanisms may drive the increase in RS scatter towards lower masses. 
The narrowness of the RS was first observed for bright galaxies; as most all galaxies even in our highest mass bin were dimmer than $.4 \, L_*$, 
it should be relatively unsurprising that our findings in lower mass bins run contrary to high-mass expectations. 


\begin{figure}\centering
  \includegraphics[width=\linewidth]{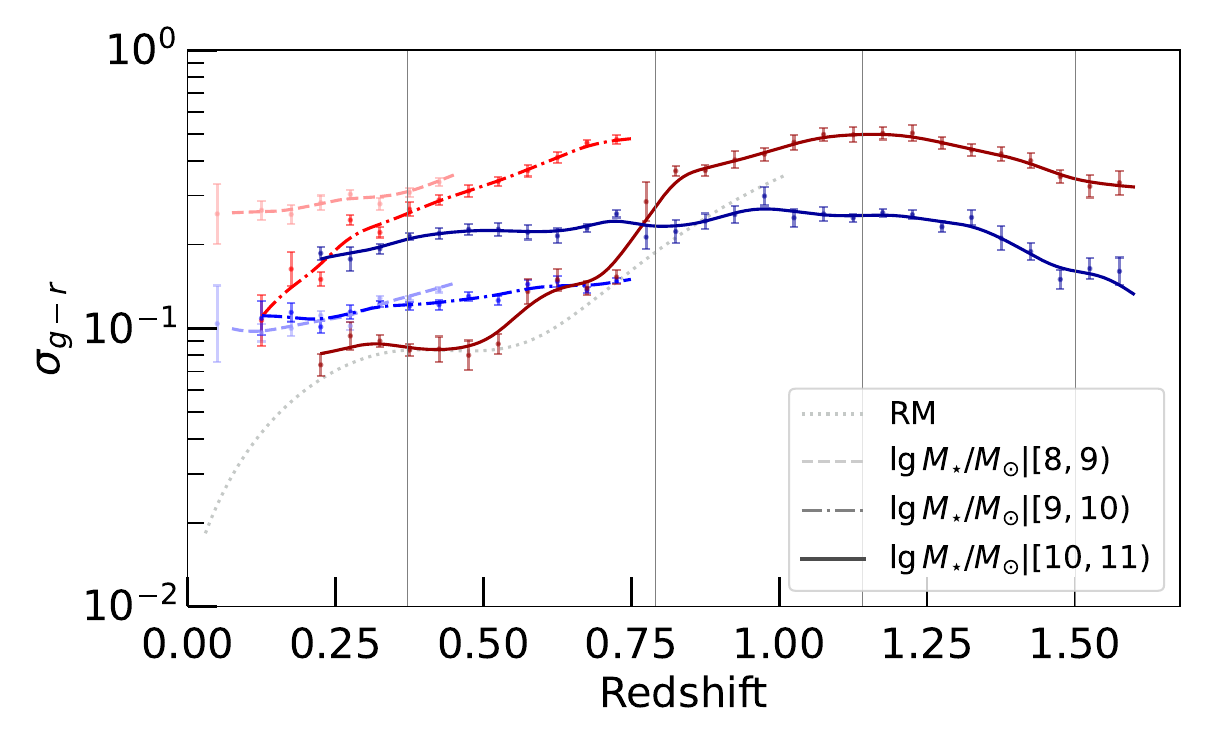}
  \vspace{-16 truept} 
  \caption{
    As Figure~\ref{fig:griz_sig}, but for $g-r$. 
  }
  \label{fig:griz_sig_gmr}
\end{figure}

\begin{figure}\centering
  \includegraphics[width=\linewidth]{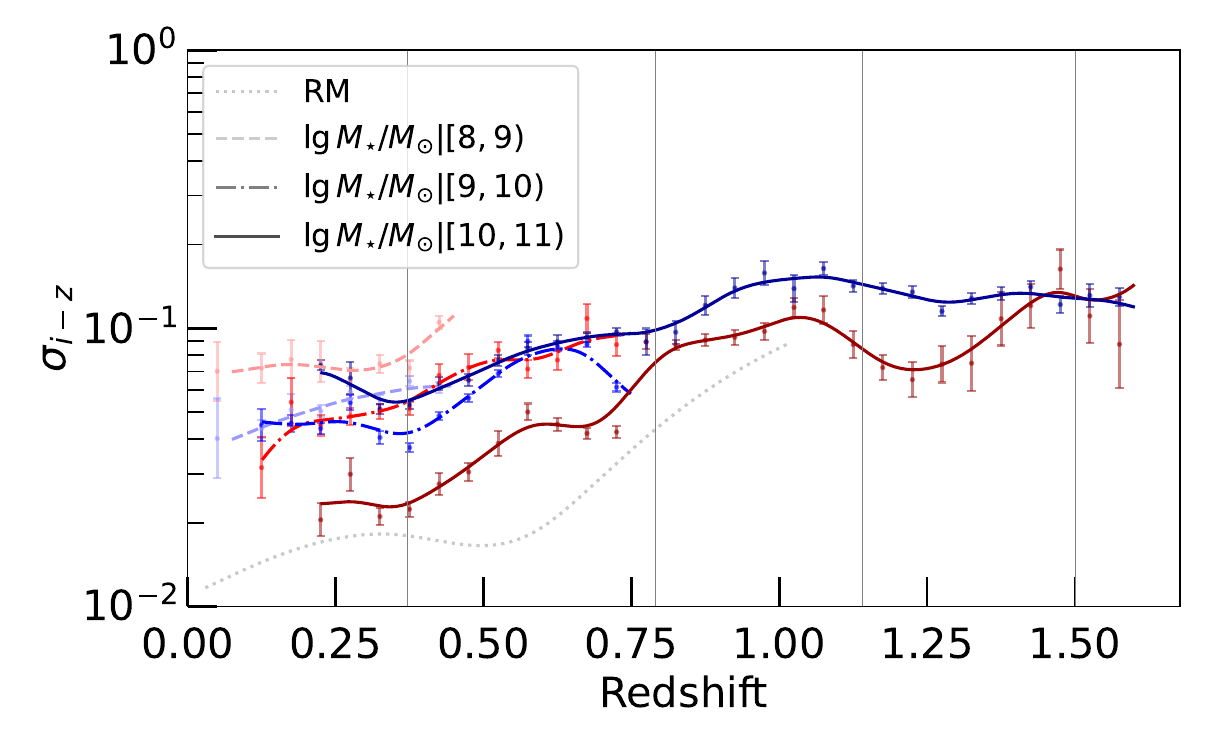}
  \vspace{-16 truept} 
  \caption{
    As Figure~\ref{fig:griz_sig}, but for $i-z$. 
  }
  \label{fig:griz_sig_imz}
\end{figure}

Figure~\ref{fig:griz_sig_imz} shows RD measuring a significantly larger RS $i-z$ scatter than RM, by roughly a factor of three in places (most often closer to a factor of two). 
The RM sample focuses on cluster members whereas the COSMOS sample focuses on field members; it could be that intrinsic color scatter truly differs between the two samples, with cluster members exhibiting less variation in near-infrared slope than field members. 

\subsection{Intrinsic color correlations}
In this section, we discuss correlations between photometric colors for RS \& BC. 
Perfect correlation ($\rho = 1$) implies as one color reddens, the other is guaranteed to as well. Power-law spectral divergences have such an effect; e.g. variations in dust content uniformly redden (or bluen) the entire spectrum slope in the optical regime. 
Negative correlation implies as one color reddens, the other bluens. Strong variations in line emission have such an affect in overlapping\footnote{
    Here I define `overlapping' correlations as those where the colors which share a mutual photometric band, such as $\rho(r-i, \, i-z)$ sharing $i$-band. 
} color correlations: if $i$-band measures some strong variable line feature, then $r-i$ will increase while $i-z$ will decrease (and vice versa), causing a negative correlation. 


\begin{figure}\centering
  \includegraphics[width=\linewidth]{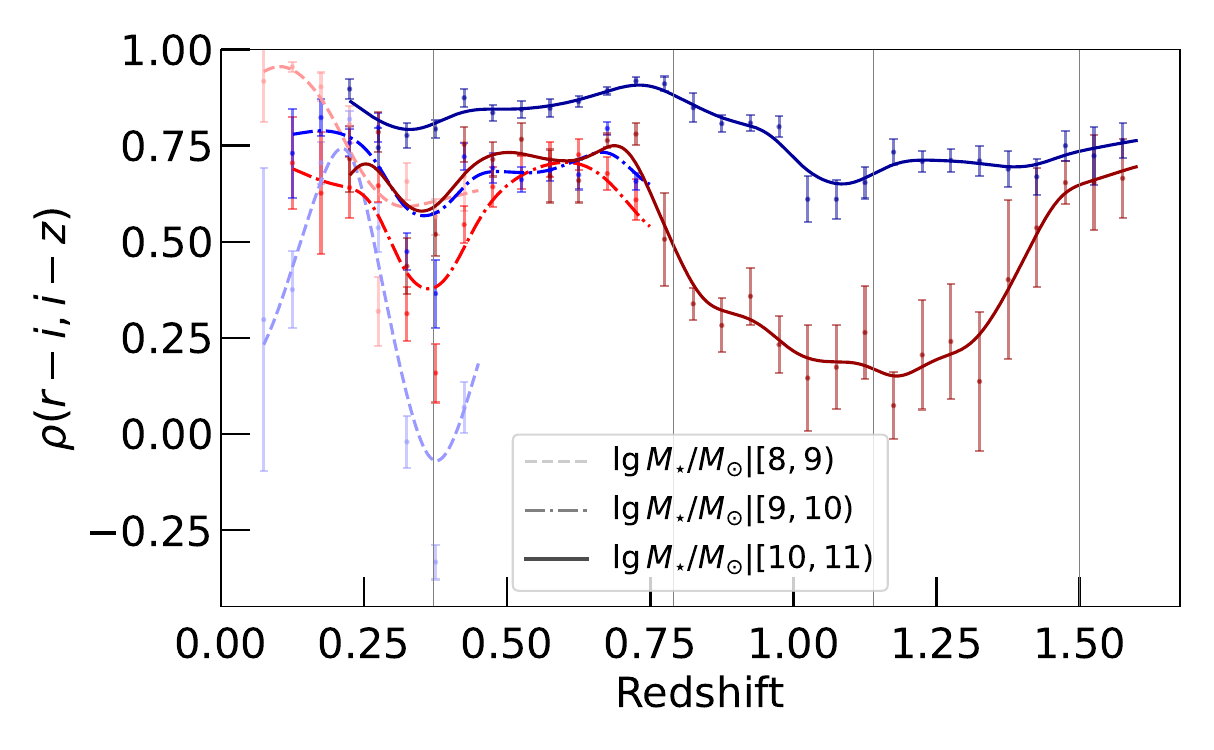}
  \vspace{-16 truept} 
  \caption[COSMOS inter-color correlations, $\rho(r-i, \, i-z)$]{
    RD-measured inter-color correlation between photometric colors $r-i$ and $i-z$; lines as in Figure~\ref{fig:griz_mu}. 
    See Figures~\ref{fig:griz_corr_gmr_rmi} \&~\ref{fig:griz_corr_gmr_imz} for characterizations of $\rho(g-r, \, r-i)$ \& $\rho(g-r, \, i-z)$. 
  }
  \label{fig:griz_rho}
\end{figure}

Figure~\ref{fig:griz_rho} shows redshift evolution of intrinsic inter-color correlations between the overlapping colors $r-i$ and $i-z$ for each mass bin and for each component as they evolve across redshift. 
%
Broadly speaking, correlations tend to decrease over redshift ($d\rho/dz \sim -.16$) and increase with stellar mass ($d\rho/d\lg M_\star / M_\odot \sim +.12$). The BC typically displays higher correlations than the RS ($\rho_{\rm BC} - \rho_{\rm RS} \sim .1$), particularly so at higher masses. 
%
Very roughly, we find inter-color correlation for the RS is $\rho_{\rm RS} \sim .7$ whereas the BC has typical correlation $\rho_{\rm BC} \sim .8$. 

High correlations $\left| \rho \right| \sim 1$ indicate that the two colors are largely redundant (i.e. adding the second color or removing it from the model won't drastically change parameterization) whereas low correlations $\left| \rho \right| \sim 0$ indicate that the two colors give more complementary information. 
As we generally find $\rho_{\rm BC} > \rho_{\rm RS}$ (particularly so at higher masses), a single color characterizes BC galaxies better than a single color characterizes RS galaxies; having more photometry greatly aids in RS selection, but doesn't aid BC selection to the same extent.

\begin{figure}\centering
  \includegraphics[width=\linewidth]{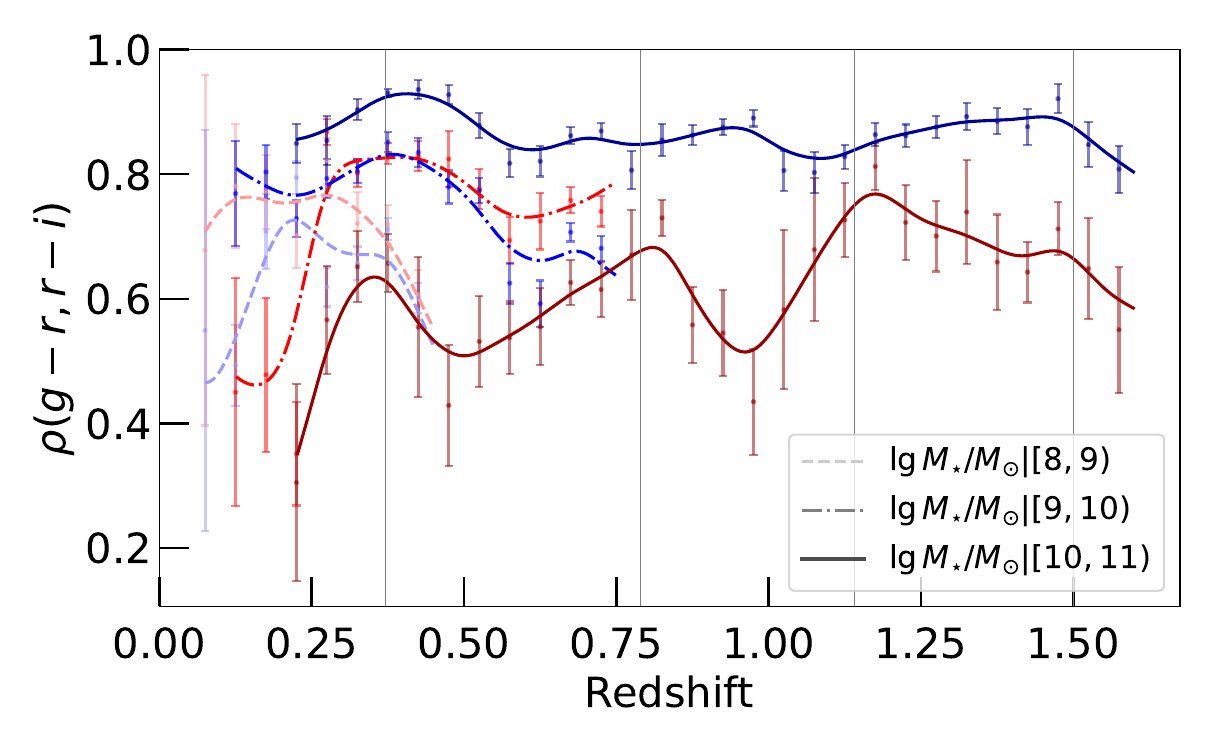}
  \vspace{-16 truept} 
  \caption{
    As Figure~\ref{fig:griz_rho}, but for $\rho(g-r,r-i)$. 
  }
  \label{fig:griz_corr_gmr_rmi}
\end{figure}

\begin{figure}\centering
  \includegraphics[width=\linewidth]{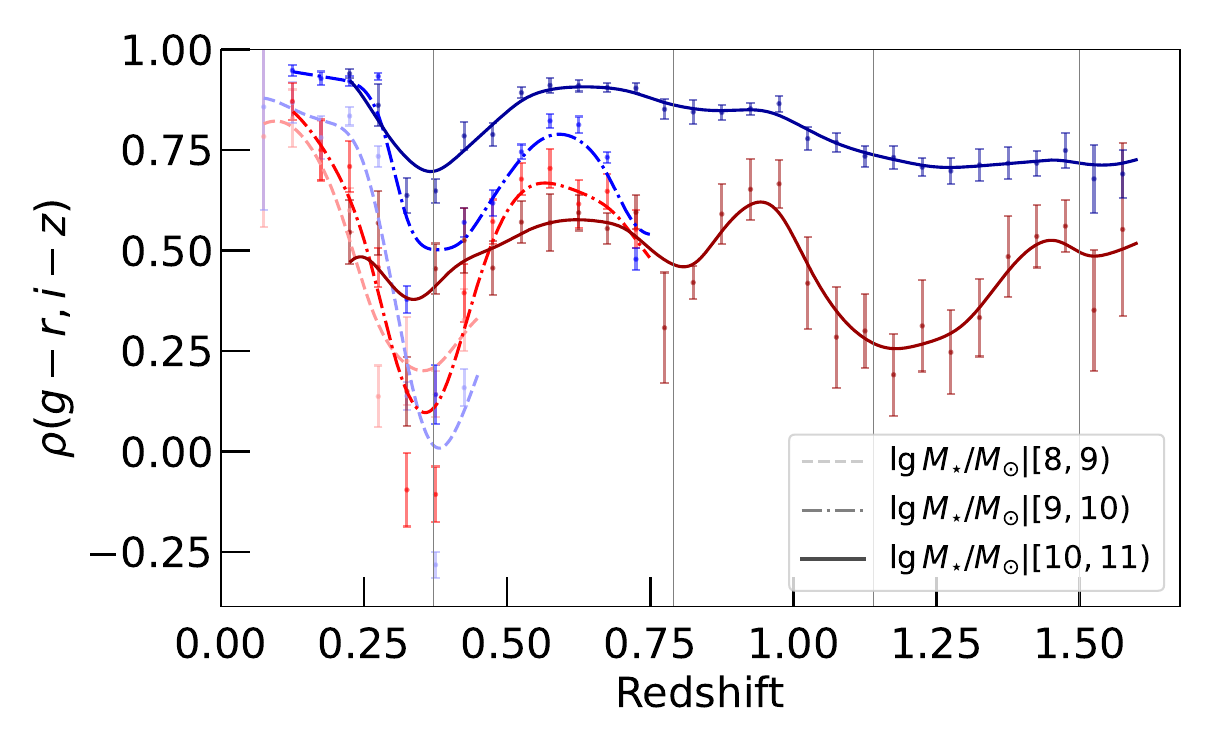}
  \vspace{-16 truept} 
  \caption{
    As Figure~\ref{fig:griz_rho}, but for $\rho(g-r,i-z)$. 
  }
  \label{fig:griz_corr_gmr_imz}
\end{figure}

Figure~\ref{fig:griz_corr_gmr_imz} shows a dip around $z=.37$ as Figure~\ref{fig:griz_rho}, though the same does not manifest for Figure~\ref{fig:griz_corr_gmr_rmi}; the feature only manifests for those correlations which involve the color $i-z$. 
  At that redshift, $i-z$ centers about $\lambda_{\rm rest} \sim 6200~\Angstrom$, not far from the H$\alpha$ line (at $6550~\Angstrom$). 
  (Alternatively, a feature captured by $z$-band alone could explain this peak. Due to atmospheric absorption, $z$-band exhibits a dual-peaked transmission function, with its maximum transmission peak near $9000~\Angstrom$. At redshift $z \sim .37$, $z$-band centers about $\lambda_{\rm rest} \sim 6600~\Angstrom$, right at the H$\alpha$ emission line.) 
Further research can discern the exact cause of this dip in correlation.

While scatter in BC colors is primarily caused by a single factor (variation in dust content), scatter in RS colors is driven by a multiplicity of factors (age, metallicity, sSFR, and to a lesser extent dust). 
Because dust affects the entire optical spectrum similarly (with Rayleigh scattering driving flux divergences roughly $\propto \lambda^{-4}$), variations in BC colors highly correlate. 
In contrast, the many causes of variation in RS spectra (e.g. age vs sSFR) affect colors non-uniformly, resulting in lower correlations. 
While there is only one galactic main sequence (BC), many different quenching pathways exist to move galaxies from the BC to the RS, be it by aging or feedback or merging. These various paths imprint themselves differently on RS spectra, again resulting in lower correlations between RS galaxies than between BC galaxies.

Future research can reveal the causes of the particular features in correlation shown above. 
Careful SPS analysis could lay down theory expectations for astrophysical interpretation of these correlations; confirmation and further interpretation of features could come from spectroscopic datasets, such as from the Dark Energy Spectroscopic Instrument \citep[DESI; see][]{DESI_Collaboration_2016}. 
We leave deeper analysis of inter-color correlation features to future papers.


\section{Rest-frame photometric color} \label{sec:color_interp}

\iffalse
  Figure~\ref{fig:lam_rest} displays 
\else
  Here we display 
\fi 
colors as functions of rest-frame wavelength, as imputed from transition redshifts (of Appendix~\ref{apx:Z_transition}). As mentioned in \S\ref{sec:color_scatt}, filters have substantial width and asymmetries, leading to these color measurements being smoothed and distorted compared to `true' (instantaneous) measurements, using infinitesimally narrow bandpasses. (For example, $g$-band is roughly a right triangle while $i$-band is roughly a box function; a pure white signal would be captured as $+.6~{\rm mag}$ dimmer in $g$-band than in $i$-band.) 
This means that (approximate) spectral slopes measured by different photometric colors may be somewhat offset from each other, despite measuring the same rest wavelength range (even if the spectrum is unchanging over time). This makes comparison of vertical offsets between different colors generally less profitable than comparing relative shapes between colors (e.g. alignment of peaks) or same-color differences between mass bins (e.g. how peaks grow with mass).

\begin{figure*}\centering
  \includegraphics [width=.49\linewidth] {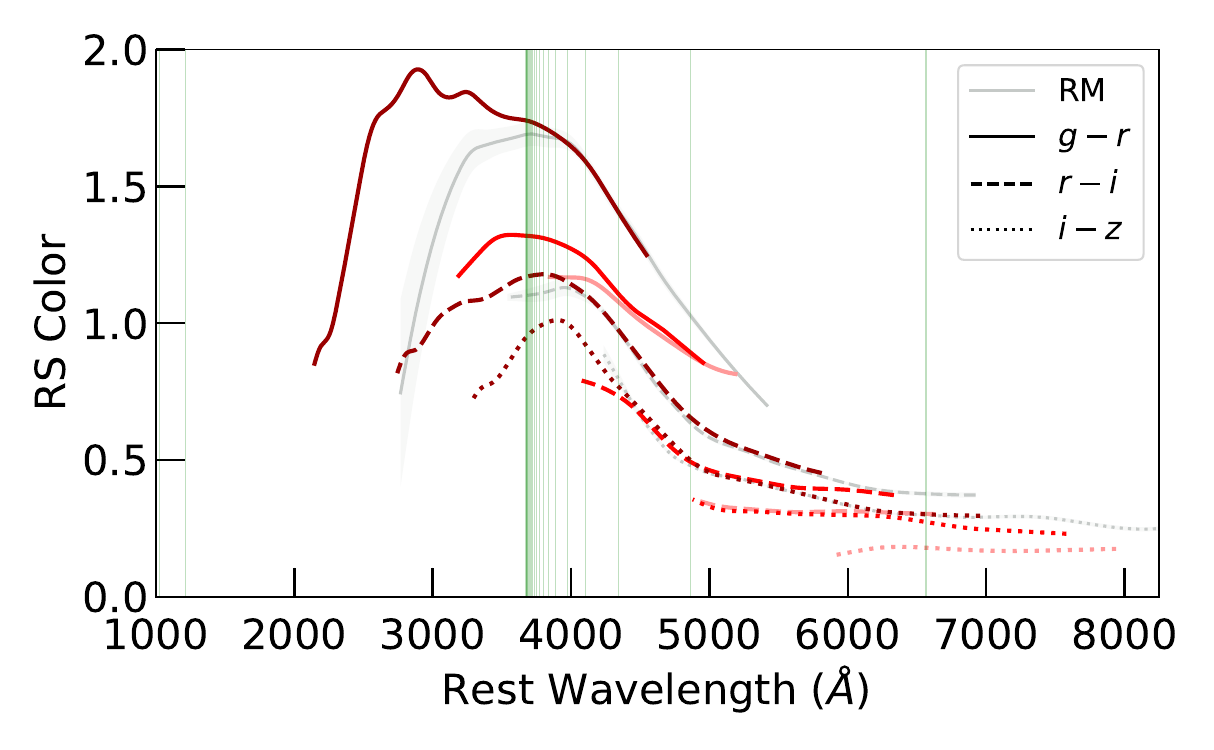} 
  \includegraphics [width=.49\linewidth] {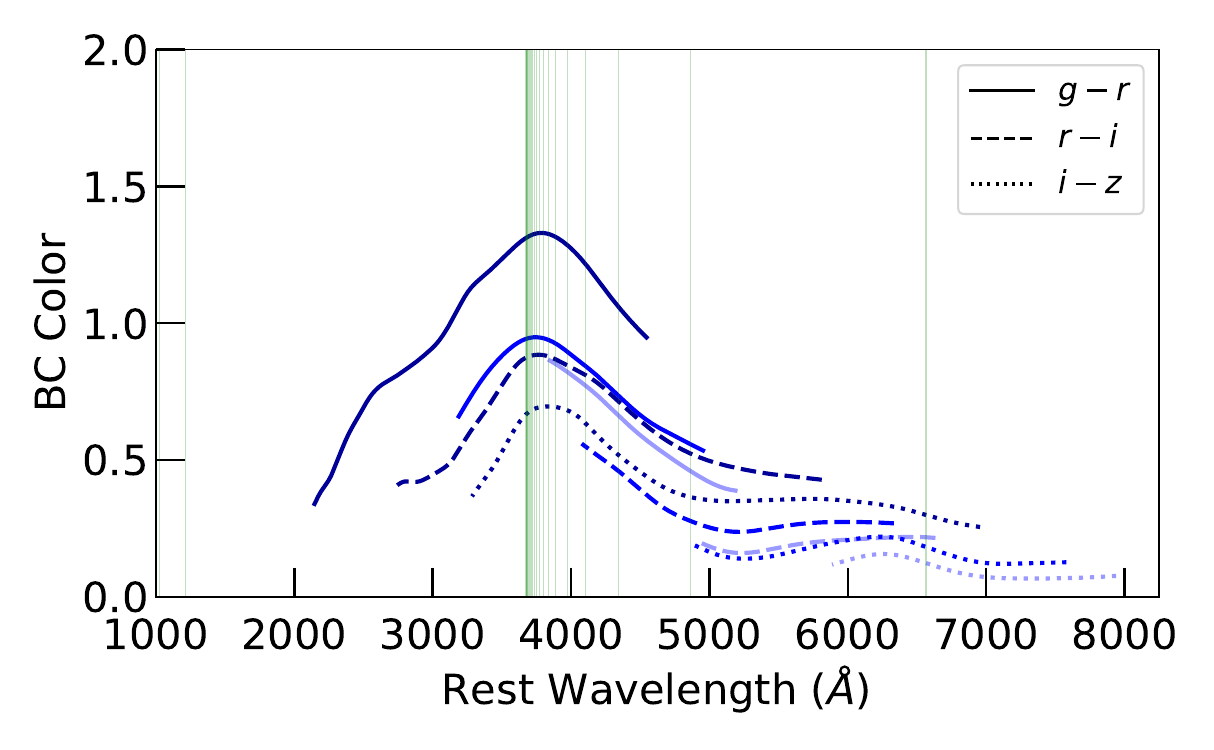} 
  \vspace{-8 truept} 
  \caption[Mean colors vs imputed rest-frame wavelength]{
    As Figure~\ref{fig:scat_rest}, but for mean colors. 
  }
  \label{fig:lam_rest}
\end{figure*}

Figure~\ref{fig:lam_rest} shows measured mean colors transformed from their measured domain of redshift (as shown in Figures~\ref{fig:griz_mu}, \ref{fig:griz_mu_gmr}, \& \ref{fig:griz_mu_imz}) to an imputed rest-frame wavelength $\lambda_{\rm rest}$. 
The most notable feature for both RS \& BC is the major peak, aligning near wavelength $\lambda \sim 3900~\Angstrom$ (corresponding to the major RS and BC peaks seen in Figure~\ref{fig:griz_mu} above, at $z \sim .9$ for $r-i$). 
In addition, a significant second peak emerges for BC galaxies near $\lambda = 6300~\Angstrom$ (the small BC bump seen at $z \sim .13$ {\it ibid.}). Several tertiary features are also present, with unresolved peaks at wavelengths such as near $2500~\Angstrom$ in the BC. 

As mentioned earlier, a step function in flux would register as a single triangle pulse in neighboring photometric colors. This roughly explains the primary peak, stepping up near $\lambda = 3900~\Angstrom$ for both RS and BC (to a lesser extent), caused by the combination of metallic lines and the Balmer series termination (at $\lambda \geq \lambda_{\rm B} \doteq 3647~\Angstrom$). 

A narrow Gaussian function in flux (like an emission line) would register in neighboring photometric colors as a tilde-like shape, with its center corresponding to the peak of the pulse. This roughly explains the secondary peak near $\lambda = 6300~\Angstrom$ (and the accompanying dip near $6800~\Angstrom$) in the BC: the midway inflection points across all colors and masses roughly correspond to the H-$\alpha$ line (at $\lambda \doteq 6550~\Angstrom$) of the hydrogen spectral series, coincident with [N\,{\sc ii}] and [S\,{\sc ii}] emission. 
This agrees with expectations, as the H$\alpha$ peak is a major spectral feature of star-forming galaxies \citep{Kriek+11}. 

Tertiary peaks hint at other features, such as the unresolved RS $g-r$ peak near $2800~\Angstrom$ (and perhaps in the BC a tilde-like feature about the same wavelength). 
We forbear from analyzing these features in this paper. 

While some differences between colors may be due to filter inhomogeneities, it may be that the differences are driven by redshift evolution. For example, around $3250~\Angstrom$ in the RS plot, not only are different bands and masses offset vertically (which could easily be caused by asymmetries in filter widths), but the curves show distinct shapes, with the $\langle g-r \rangle$ curves remaining redder than other colors (which, at the same imputed rest wavelength, are at higher redshift). This could imply a significant reddening of spectra (steepening of spectral slope) about $3250~\Angstrom$ over cosmic time.


Despite being only a rough first pass, our imputation of rest-frame spectral slopes reveals significant spectral differences between RS and BC on average; in addition to the $4000~\Angstrom$ break, we detect H$\alpha$ emission in the BC. 
Future analyses could use more detailed methods to align slopes between bands, accounting for differences between bandpass transmissions. Accounting for these asymmetries and deformities in bandpass shape (as compared to flawless box functions) would allow for more accurate spectrum reconstruction, improving characterization of populations. This could better decipher whether features are merely caused by band asymmetries or by redshift evolution.


\section{Permissible bin thickness} \label{sec:permissible_width}
What step size in redshift or galactic stellar mass causes a significant drift in RD fit parameters $\vec\theta$? 


\subsection{Color slope method}

One straightforward way to quantify permissible bin widths is to compare how much the RS mean color drifts in comparison to its scatter. Using redshift as an example (though one can mirror the analysis using stellar mass), if a step $\Delta z$ in redshift makes the mean color shift by significantly more than its scatter, we must consider decreasing redshift bin width in order to not miss significant spectral features. 
As $\pm 2 \sigma_{\rm RS}$ has historically been used to select the RS, if the mean color drifts by $>4 \sigma_{\rm RS}$ across a given bin, the color would then be completely excluded from selection. We can then demand the bin width $\Delta z$ times the redshift-derivative of color $d\vec c_{\rm RS} / dz$ remain smaller than $4\sigma_{\rm RS}$, so 
\begin{align}
  \Delta z & < \frac{4 \sigma_{\rm RS}}{d\vec c_{\rm RS} / dz} = \Delta z_{\max}. 
\end{align}
This then estimates a maximum $z$-bin width, beyond which colors vary more within the bin than they scatter at a fixed $z$. A more conservative bin width would be half this value, but there would be little justification to move bin width below a quarter this value (unless, as will be discussed in \S\ref{sec:f_xx}, other parameters drift significantly quicker than mean color does compared to its scatter). Across all colors considered, the minimum value of $\Delta z_{\max}$ would then be the strictest requirement on permissible bin size. 

From our three main fits here, we find $\Delta z_{\max}$ tends to be lowest (most strict) at lower redshifts and at higher stellar masses. For our samples, we measure a minimum of $\Delta z_{\max} \sim .12$, implying that redshift bins should not exceed this width if the RS is to be properly characterized (at low redshifts and high stellar masses). At lower stellar masses, larger RS scatter $\sigma_{\rm RS}$ and smaller RS redshift slope $d \vec c_{\rm RS} / dz$ lead to higher values of $\Delta z_{\max}$; we find values of .25 and .75 for the middle and lowest mass sample respectively. 
(This analysis can also be performed for the BC; we find strictest values of $\Delta z_{\max}$ for each sample were consistent with $\sim .3$). 

This analysis can also be performed with dragons which run with stellar mass, using $d \vec c / d\lg M_\star / M_\odot$ rather than $d \vec c_{\rm RS} / dz$. Preliminary results find strictest values of $\Delta \lg M_{\star, \max} / M_\odot \sim 1$, implying that within a stellar mass decade, the RS and BC mean colors tend to not experience a drift in mean colors greater than four times the RS scatter. 
However, parameters besides mean color may experience significant drift with redshift or stellar mass, suggesting stricter bin width requirements.

\subsection{Parameter curvature method} \label{sec:f_xx}

A more conservative approach could use bootstrap uncertainties of fit parameters along with curvature of fit parameters to determine desired bin width. 

Generally speaking, if an unknown function $f(x)$ is linear in a given domain, then only two points (at any distance from each other) are needed to characterize the curve, with no constraint on bin width; in contrast, if the function has significant curvature, this constrains bin size. In particular, one might hope to have sufficiently sampled space so as to properly characterize all curvatures in the function, ensuring that all peaks are identified. If all points sampled of the function have some uncertainty, then one could demand \emph{insignificant curvature between neighboring points}. This would distinguish outliers from significant peaks. 

In particular, using finite differencing, the curvature $f_{xx}$ of a function at point $x_i$ is estimated by 
\begin{equation}
  f_{xx}(x_i) \approx \frac{f_{i-1} - 2 f_{i} + f_{i+1}}{h^2}
\end{equation}
(where $h$ is bin width). Demanding no significant curvature between points would then imply $f_{xx} / \sigma_{f_{xx}} \lesssim 1$. 
We can then estimate an optimal bin width using this constraint. Because $f_{xx} \propto h^{-2}$, in order to make point-to-point curvature insignificant, we must use bin width 
\begin{equation} \label{eqn:h_new}
  h_{\rm new} = h_{\rm old} \cdot \sqrt{ \frac{ \sigma_{f_{xx}} }{ f_{xx} } }. 
\end{equation}
Thus, larger relative uncertainties on curvature $\sigma_{f_{xx}}/f_{xx}$ allow for larger bin widths while more significant curvatures $f_{xx}/\sigma_{f_{xx}}$ demand narrower bins. 
Using $h_{\rm new}$ thus ensures that all neighboring points are consistent to $1 \sigma$ with linear, attesting that the function is well-characterized, without missing any significant peaks of $f(x)$. 

Applying equation~\eqref{eqn:h_new} to parameters $\vec\theta$ from our dragons (with accompanying uncertainties) then gives us a strict estimate on bin width, setting a minimum reasonable bin width for a given dragon. Rather than demand $f_{xx} / \sigma_{f_{xx}} \lesssim 1$, one could reasonably allow for significance of curvature up to $f_{xx} / \sigma_{f_{xx}} \lesssim 5$, only excluding undeniably significant curvatures. Restating these bounds in terms of a given parameter $\theta$, we then have 
\begin{equation}
  h_{\rm old} \cdot \sqrt{ \frac{\sigma_{\theta_{xx}}} {\theta_{xx}} }
  \lesssim h_{\rm new} \lesssim 
  h_{\rm old} \cdot \sqrt{ \frac{5 \sigma_{\theta_{xx}}} {\theta_{xx}} }.
\end{equation}
Using the left constraint for component weights, mean colors, log variances, and correlations, we find a lower bound of $h_{\rm new}$ for redshift bins around $\Delta z = .04$; the minimum across all parameters and both components is $\Delta z = .013$, yielding an upper constraint of $\Delta z \lesssim .03$ as a strict value of a maximum permissible bin width for redshift. Our study's bin width of $\Delta z = .05$ is thus slightly large (as was desired for the sake of increasing RS number counts in each bin, improving statistical power). 

Constraints on bin width are often most stringent due to sharp evolution of colors, but depending on the sample, other elements of $\vec\theta$ demand narrower bin widths: weight, correlation, or even scatter (in decreasing occurrence) can also sound the strictest calls for narrow bin widths.

The same methodology can be used for magnitude-running dragons. Analyzing one such dragon, we find preliminary results of a typical desired bin width of $\Delta \lg M_\star / M_\odot \sim .35$ and a minimum desired bin width of $\Delta \lg M_\star / M_\odot > .1$ (implying an upper limit of $\Delta \lg M_\star / M_\odot \lesssim .22$ for negligible parameter evolution within a mass bin). 
Using bins any thinner than a tenth of a mass decade is therefore unreasonable for the purposes of GMM characterization of galaxy colors. Furthermore, using a full decade for stellar mass bin width is too lenient, as it misses significant parameter evolution. 

\subsection{Suggested bin widths for redshift \& mass}

We find no evidence of needing bin width thinner than $\Delta z < .01$ for redshift nor $\Delta \lg M_\star / M_\odot < .1$ for stellar mass. 
Future modeling of galaxy populations in photometric space should ensure their models use bin widths $\Delta z \lesssim .12$ and $\Delta \lg M_\star / M_\odot \lesssim 1$ in order to properly model colors. 
To properly model the complete set of GMM fit parameters, bin widths should be further narrowed to $\Delta z \lesssim .03$ and $\Delta \lg M_\star / M_\odot \lesssim .22$, otherwise GMM fitting risks missing significant evolution of parameters within their bins. 
While using the former, larger maximum bin widths will improve statistical power, this reduces parameter interpretability.

If mass bins are made too wide, then high-$M_\star$ BC galaxies are likely to be characterized as RS galaxies. As this population tends to have lower star formation rates and is likely in the process of mass quenching, this characterization may not be entirely incorrect from an sSFR perspective, though it is perhaps wrong from a complete astrophysical perspective, as it may e.g. model dusty spirals as RS members.








\onecolumngrid

\end{document}